\documentclass[a4paper,preprintnumbers,floatfix,superscriptaddress,pra,onecolumn,showpacs,notitlepage,longbibliography,unpublished]{quantumarticle}
\pdfoutput=1
\usepackage{amsmath,amsthm,amsfonts,amssymb,mathtools,dsfont,mathrsfs}
\usepackage{bm,bbm}
\usepackage{cancel}
\usepackage{marvosym}
\usepackage{graphicx}
\usepackage{xcolor}
\usepackage{xr}
\usepackage{float}
\usepackage{appendix}
\usepackage{graphicx}
\usepackage{dcolumn}
\usepackage{xcolor}
\usepackage{scalerel}
\usepackage{soul}
\usepackage{tikz}

\usepackage{mathdots}
\def\myvdots{\ \vdots\ } 

\def\myidots{\ \iddots\ }
\usepackage{stmaryrd}
\usepackage{svg}
\usepackage{hyperref}
\PassOptionsToPackage{linktocpage}{hyperref}
\usepackage[capitalize]{cleveref}
\crefname{figure}{Fig.}{Figs.}
\crefname{algorithm}{Protocol}{Protocols}

\usepackage{crossreftools}
\newcommand{\optionaldesc}[2]{%
  \phantomsection
  #1\protected@edef\@currentlabel{#1}\label{#2}%
}

\usepackage{acronym}
\usepackage{algorithm}
\usepackage[noend]{algorithmic}
\floatname{algorithm}{Protocol}

\DeclareMathOperator{\GHZ}{GHZ}
\usepackage{tikz-cd}

\makeatletter
\newtheorem*{rep@theorem}{\rep@title}
\newcommand{\newreptheorem}[2]{%
\newenvironment{rep#1}[1]{%
 \def\rep@title{#2 \ref{##1}}%
 \begin{rep@theorem}}%
 {\end{rep@theorem}}}
\makeatother
\newtheorem{theorem}{Theorem}
\newreptheorem{theorem}{Theorem}
\newtheorem{result}[theorem]{Result}

\newtheorem{lemma}[theorem]{Lemma}
\newreptheorem{lemma}{Lemma}
\newtheorem{proposition}[theorem]{Proposition}
\newreptheorem{proposition}{Proposition}
\newtheorem{definition}[theorem]{Definition}
\newtheorem{remark}[theorem]{Remark}
\newtheorem{corollary}[theorem]{Corollary}
\newtheorem{techlemma}{Lemma}[section]

\newtheorem{task}{Task}

\DeclarePairedDelimiter{\bra}{\langle}{\vert}
\DeclarePairedDelimiter{\ket}{\vert}{\rangle}
\DeclarePairedDelimiterX\braket[2]{\langle}{\rangle}%
  {#1\kern0.15ex\delimsize\vert\kern0.15ex\mathopen{}#2}

\DeclarePairedDelimiterX\ketbra[2]{\vert}{\vert}%
  {#1\kern0.15ex\delimsize\rangle\delimsize\langle\kern0.15ex\mathopen{}#2}

\DeclarePairedDelimiterX{\abs}[1]{\lvert}{\rvert}{%
  \ifblank{#1}{\,\cdot\,}{#1}
}   
\DeclarePairedDelimiterX\norm[1]\lVert\rVert{%
  \ifblank{#1}{\,\cdot\,}{#1}
}   

\newcommand{\kb}[2]{\ketbra{#1}{#2}}
\newcommand{\avg}[1]{\langle #1\rangle}
\renewcommand{\norm}[1]{\left\lVert#1\right\rVert}
\renewcommand{\abs}[1]{\ensuremath \left|#1\right|}

\newcommand{\id}{\mathrm{id}}
\newcommand{\CC}{\mathbb{C}}

\newcommand{\ZZ}{\mathbb{Z}}
\newcommand{\NN}{\mathbb{N}}

\newcommand{\WW}{\mathcal{W}}
\DeclareMathOperator{\EE}{\mathbb{E}}

\newcommand{\ii}{\ensuremath\mathrm{i}}

\newcommand{\veps}{\varepsilon}

\newcommand{\obda}{without loss of generality }

\DeclareMathOperator{\tor}{tor}

\DeclarePairedDelimiterX\Set[1]\{\}{%
   
#1
}

\renewcommand{\vec}[1]{\mathbf{#1}}

\DeclareMathOperator{\supp}{supp}

\DeclareMathOperator{\tr}{Tr}
\DeclareMathOperator{\poly}{poly}

\DeclareMathOperator{\argmax}{argmax}
\DeclareMathOperator{\sgn}{sgn}

\usepackage{quantikz}
\usetikzlibrary{quantikz2,fit,backgrounds,positioning,calc}

\newacro{POVM}{positive operator-valued measure}
\newacro{PVM}{projection-valued measure}
\newacro{CP}{completely positive}
\newacro{CPTP}{completely positive trace preserving}
\newacro{PSD}{positive semidefinite}
\newacro{NISQ}{noisy intermediate-scale quantum}
\newacro{ONB}{orthonormal basis}
\newacro{PCA}{principal component analysis}
\newacro{WH}{Weyl-Heisenberg}
\newacro{MUB}{mutually unbiased bases}
\begin{document}
\title{\texorpdfstring{An infinite hierarchy of multi-copy\newline quantum learning tasks}{An infinite hierarchy of multi-copy quantum learning tasks}}

\author{Jan N\"oller}
\affiliation{Quantum Computing Group, Department of Computer Science, Technical University of Darmstadt, Germany}
\email{jan.noeller@tu-darmstadt.de}
\author{Viet T. Tran}
\affiliation{Department for Quantum Information and Computation at Kepler (QUICK), Johannes Kepler University Linz, Austria}
\author{Mariami Gachechiladze}
\affiliation{Quantum Computing Group, Department of Computer Science, Technical University of Darmstadt, Germany}
\author{Richard Kueng}
\affiliation{Department for Quantum Information and Computation at Kepler (QUICK), Johannes Kepler University Linz, Austria}
\date{}

\begin{abstract} 
Learning properties of quantum states from measurement data is a fundamental challenge in quantum information. The sample complexity of such tasks depends crucially on the measurement primitive. While shadow tomography achieves sample-efficient learning by allowing entangling measurements across many copies, it requires prohibitively deep circuits. At the other extreme, two-copy measurements already yield exponential advantages over single-copy strategies in tasks such as Pauli tomography. In this work we show that such sharp separations extend far beyond the two-copy regime: for all integer that are not divisible by four, we construct explicit learning tasks of \textit{degree} $c$, which are exponentially hard with $(c-1)$-copy measurements but efficiently solvable with $c$-copy measurements. Importantly, our hardness results do not hinge on any assumptions about precision scaling. Moreover, our protocols are not only sample-efficient but also realizable with shallow circuits. 
Together, our results reveal an infinite hierarchy of multi-copy learning problems, uncover new phase transitions in sample complexity, and underscore the role of reliable quantum memory as a key resource for exponential quantum advantage.
\end{abstract}

\maketitle

\hypersetup{pdftitle = {An infinite hierarchy of multi-copy quantum learning tasks},
       pdfauthor = {Jan Nöller, Viet T. Tran, Mariami Gachechiladze, Richard Kueng},
       pdfsubject = {Quantum computing}
      }

\section*{Introduction}
Quantum state learning is the fundamental task of inferring useful information about an unknown quantum system from experimental data. The data are obtained through quantum measurements, which inevitably disturb the system. To collect sufficient statistics, the same quantum state must therefore be re-prepared many times, with each copy measured individually. The total number of preparations and measurements needed to estimate the desired properties with high confidence is known as the \emph{sample complexity}. This parameter is central in determining the feasibility of quantum learning protocols.  

The most general approach to state learning is \emph{quantum state tomography}~\cite{cramer2010efficient}, which reconstructs the full density matrix of the system. For an $n$-partite quantum system, the density matrix is a Hermitian $d^n \times d^n$ matrix in a local dimension $d$, and tomography corresponds to accessing all of its informationally complete degrees of freedom. While this procedure yields complete knowledge of the state, it comes with a prohibitive cost: even the best-known tomographic techniques~\cite{krt2017,haah2016sample,ow2016,gkkt18} require a number of samples that grows exponentially with $n$. This exponential overhead is unavoidable, as shown by fundamental lower bounds~\cite{haah2016sample,nayak2025lower}.  

In practice, however, one is rarely interested in reconstructing the entire density matrix. More often, the goal is to predict a family of properties of the state. This motivates the framework of \emph{shadow tomography}~\cite{aaronson2020shadow}, where the task is to approximate the expectation values of many observables $\mathrm{tr}(O_l \rho)$ up to accuracy $\epsilon$, without fully reconstructing $\rho$. For $L$ observables on an $n$-qubit system, refined shadow tomography protocols achieve this with a sample complexity of $\widetilde{O}(n \log^2(L) / \epsilon^4)$ copies \cite{badescu2024improved}, remarkably efficient and independent of both the state and the observables. The trade-off, however, is that the required quantum circuits are extremely demanding: they involve exponentially deep operations acting on all of these state copies simultaneously.  

A natural way to reduce these overheads is to restrict either the class of states or the class of observables. \emph{Classical shadows}~\cite{huang2020predicting,morris2019selective,paini2019approximate,elben2023randomized} are a prominent example of the latter. A particularly important subclass of observables is given by the $n$-qubit Pauli operators, i.e., tensor products of single-qubit Pauli matrices. These play a central role in quantum information, quantum algorithms, and qubit encodings of quantum chemistry problems; see e.g.~\cite{huang2021derandomization}. Learning Pauli observables from data is commonly referred to as \emph{Pauli shadow tomography}, and because it relies on sequential single-copy measurements, it requires exponentially many samples in the worst case~\cite{huang2021information,huang2022quantum}.  

With the continued development of quantum hardware, these no-go results can be circumvented by changing the measurement paradigm: instead of measuring single copies in a sequential fashion, one may store multiple copies in a quantum memory and process them jointly using entangling measurements. Refs.~\cite{huang2021information,huang2022quantum} have shown that this drastically changes the picture: With access to only two copies of a quantum state at a time and the ability to perform entangling measurements on both copies, it becomes possible to jointly estimate the absolute values of all $4^n$ squared Pauli expectation values, $\lvert \tr(P\rho) \rvert^2$, in a sample-efficient manner. What is more, the quantum circuits involved only have constant depth.  When combined with lower bounds showing that the same task is exponentially costly in the single-copy regime, this establishes a rigorous exponential sample advantage~\cite{huang2021information} that has been demonstrated on a 53-qubit quantum device~\cite{huang2022quantum}.  

Beyond Pauli tomography, similar two-copy advantages have been demonstrated for principal component analysis~\cite{koczor2021the} and for quantum channel learning~\cite{liu2025bosonicchannellearning}. Follow-up works have analyzed this phenomenon in greater depth, including estimation of Pauli channels~\cite{chen2024tight,chen2022pauli,chen2025efficient}, Pauli observables~\cite{chen2024optimal}, and purity estimation~\cite{gong2024on} (the latter being the original SWAP-test task introduced in Ref.~\cite{nakazato2012measurement}). More recently, refined techniques using martingale concentration inequalities have enabled a unified treatment of lower bounds for such tasks~\cite{chen2024optimal}, which we also use as a central tool in this work.  

All known exponential advantages fall into two broad categories. Some are fully realized with two-copy measurements~\cite{huang2022quantum,chen2022pauli,chen2024tight,chen2025efficient,chen2022tight}, while others follow the paradigm of shadow tomography~\cite{aaronson2020shadow,chen2021exponetial}, requiring joint measurements on a number of copies that grows with system size. In particular, Ref.~\cite{chen2021hierarchy} introduced a hierarchy of learning tasks, each of which is provably \emph{$c$-copy-hard}, meaning that no algorithm limited to $c$ or fewer copies can solve it efficiently. The best schemes, however, are again based on shadow tomography, requiring $\poly(nc)$-copy measurements and correspondingly complex circuits. This leaves a large gap between the regimes where tasks remain provably hard and those where they become efficiently solvable.

The situation in Refs.~\cite{huang2021information,huang2022quantum} is different, as it exhibits a sharp transition from exponential to linear sample complexity already at two-copy measurements. Motivated by this, we investigate whether further examples of such sharp transitions exist. For clarity, we refer to the number of copies at which such a transition occurs as the \emph{degree} of a learning task. A task is said to be of degree $c$ if it can be solved efficiently with $c$-copy measurements, but remains provably hard for strategies with $(c-1)$ (or fewer) copies, even if one allows for adaptive protocols utilizing deep quantum circuits and/or arbitrarily expensive classical co-processing. This unifying point of view places shadow tomography at one extreme of the spectrum ($c \gg 1$) and degree-$2$ tasks, like purity testing and Pauli shadow tomography, at the other. 

Naturally, this posits the question of whether there exist tasks of degree $c$ for some $c>2$, and whether one can identify infinite families of such tasks for increasing $c$. 

Our results answer both questions positively. Specifically, we construct explicit learning tasks of degree $c$ for each $c\neq 0\mod 4$, thereby establishing an infinite hierarchy that interpolates between the previously known degree-$2$ examples and shadow tomography, which can be regarded as a task of degree~$\infty$ (in the limit of very large system sizes).
On the technical side, we adapt the results developed in Ref.~\cite{chen2024optimal} to prove that no $(c-1)$-copy protocol (whether exponentially deep or adaptive) can solve the same $\epsilon$-approximate learning problem without incurring an exponential sample complexity in the number $n$ of identical quantum systems involved. 
Moreover, at the sharp transition, we show not only that the sample complexity becomes efficient, but also that the measurement protocols can be implemented with shallow circuits of (at most) logarithmic depth in $c$ (and independent of $n$). This highlights the power of multi-copy quantum information processing. We explicitly construct the protocol and give examples of how to transpile it on a qubit-based quantum architecture.

Taken together, these results uncover a rich and previously unexplored hierarchy of quantum learning problems. Complementing the results of \cite{ye2025exponential}, they demonstrate that sharp phase transitions in sample complexity are not confined to the one-copy vs.\ two-copy setting, but occur broadly across higher-copy regimes. This perspective not only advances our theoretical understanding of the complexity of quantum learning, but also suggests new ways of exploiting multi-copy quantum resources to achieve exponential advantages and points at the importance of building reliable quantum memory.

\paragraph{Note added}
During the final preparation stages of this work, we became aware of a concurrent work, Ref.~\cite{ye2025exponential}, which addresses the same overarching question from a different vantage point. In the following, we briefly outline and explain the main conceptual and technical differences. On a high level, two avenues exist to establish complexity separations via degree-two learning tasks. The first is the linearity of quantum mechanics, leading to advantages of $2$-copy schemes if one were to investigate quantities that involve, or can be determined using, the second moments of the distributed state $\rho$, such as purity estimation or principal component analysis. The second avenue builds on Heisenberg's uncertainty principle and related anticommutation relations, which can change drastically when moving to tensor products~\cite{huang2021information}. While \cite{ye2025exponential} sets out to expand upon the former approach to establish degree-$d$ learning tasks for any integer $d$, we follow the latter to also answer the posed question affirmatively. On the more technical level, the construction in \cite{ye2025exponential} works for any integer, whereas our results rely on $d$ not being divisible by four. However, the hardness of \cite{ye2025exponential} in estimating $\tr[O\rho^d]$ with $c<d$-copy measurements crucially depends on the targeted precision $\veps$, as it is known that $\tr[\rho^{d}]$ can be efficiently determined from the values $ \{\tr[\rho^k]\}_{k=1}^{d-1}$ when the targeted precision $\veps$ is not higher that $\veps= 2(d+1)e^{-d}$~\cite{Shin2025resource}. In contrast, our hardness results do not depend on any scaling of the precision, other than the requirement that $\veps<\frac{1}{6}$. To conclude, our results nicely complement the ones in \cite{ye2025exponential} to provide an infinite hierarchy of linear learning tasks that interpolates between the previously known degree-$2$ examples and shadow tomography, which can be regarded as a task of degree~$\infty$.\\
\noindent In an earlier version of this paper, we discussed a potential obstruction to extending the hardness results to composite integers $d$. Further analysis, motivated in part by the extension to square-free degrees, showed that this obstruction does not apply. The present version therefore revises that discussion and strengthens the construction to include odd prime powers, as established in~\cref{prop:prime_power_result}.

\paragraph{Outline}
The remainder of this manuscript is organized as follows: We begin with a few definitions to properly summarize our results afterward. We then set up preliminaries and define the learning tasks we consider in \cref{sec:preliminaries}. Following this, we first prove our hardness results in \cref{sec:lower_bound}. \Cref{sec:efficient_protocol} then continues to discuss the $d$-copy efficient scheme solving these tasks. More technical details and proofs supporting the claims in the main text, as well as details to the numerical simulations related to \cref{fig:3c_vs_1c} can be found in the Appendix.

\section{Summary of results}
In this section, we briefly introduce the notion of efficiently solving a learning task before we present the main results. 

\begin{definition}\label{def:efficient}
   Let $\mathcal{A}=\{\mathcal{A}_{n}\}_{n\in\NN}$ be a learning task for an $n$-partite quantum state $\rho$. 
We say that there exists an efficient protocol $\varepsilon$-approximately solving this task for $\varepsilon>0$ if it satisfies:  
\begin{itemize}
    \item \emph{Sample efficiency:} the protocol uses at most $O(\poly(n,1/\varepsilon))$ copies of $\rho$;
    \item \emph{Quantum efficiency:} all quantum subroutines can be implemented by circuits of size $O(\poly(n))$;
    \item \emph{Classical efficiency:} all classical postprocessing runs in $\poly(n)$ time using $\poly(n)$ memory.
\end{itemize}
A protocol that has access to measurement data on at most $c$ copies of $\rho$ per update step and meets all the criteria above is called \emph{$c$-copy efficient}.
\end{definition}

To capture the minimal complexity of learning tasks, we use the following terminology (cf.~\cite{chen2021hierarchy}).  

\begin{definition}\label{def:hard}
A learning task $\mathcal{A}$ is called \emph{$c$-copy hard} if there exists no sample-efficient protocol that solves $\mathcal{A}$ using adaptive measurements on at most $c$ copies of the state $\rho$ at a time. 
\end{definition}
Here, \emph{sample-efficient} refers only to the number of state copies used; we place no restrictions on the quantum circuit depth or the amount of classical postprocessing. Finally, combining the two notions above gives rise to the \emph{degree} of a learning task.  

\begin{definition}\label{def:degree}
A learning task $\mathcal{A}$ is said to have \emph{degree $c$} if it is $(c\!-\!1)$-copy hard, but admits a $c$-copy  efficient protocol.
\end{definition}

We now summarize our main result, which establishes a hierarchy of learning tasks of increasing degree. 
We introduce the specific tasks in the following sections, along with proofs of the corresponding complexity bounds. 

\begin{result}\label{res:main_result}
For each integer $d$ with $4\nmid d$  there exists a learning task $\mathcal{A}_\veps^{(d)}=\{\mathcal{A}^{(d)}_{n}\}_{n\in\mathbb{N}}$ of degree $d$. More precisely, for any precision parameter $0<\veps\leq\frac{1}{6}$ we have:
\begin{enumerate}
    \item[(i)] There is a $d$-copy efficient protocol which  uses $\mathcal{O}(nd\log(d)\,\varepsilon^{-2d})$ samples and quantum circuits of size $\mathcal{O}(dn)$ and depth $\mathcal{O}(d)$ to  solve $\mathcal{A}^{(d)}_{n}$ ($\veps$-approximately).
    \item[(ii)] Any protocol restricted to measurements on at most $c<d$ copies, $\veps$-approximately solving the task requires at least $N\geq \Omega\left((4/3)^nc^{-1}\veps^{-2}\exp(-c\veps)\right)$ samples of $\rho$.
\end{enumerate}
\end{result}
Importantly, and contrary to e.g.~\cite{ye2025exponential}, the hardness results in our construction do not depend on any assumption regarding the scaling of the targeted precision.
 As a short technical detour, and similarly to Refs.~\cite{huang2022quantum,chen2024optimal}, the learning tasks we study correspond to predicting the \emph{absolute values} of observables $A_1,\dots,A_N$ that form a basis of the underlying Hilbert space. As shown in part~(i) of \Cref{res:main_result}, this can be achieved $c$-copy efficiently for suitable $c$. Moreover, if one wishes to recover the full state, including phase information, our protocol can be combined with the scheme of Ref.~\cite{king2025triply}, as we explain in \Cref{sec:extension_to_full_tomography}, though at the cost of substantially higher classical and quantum computational resources. In particular, while the tasks we focus on are not full tomography, our constructions can be extended to Weyl-Heisenberg tomography under $d$-copy measurements, at the expense of forfeiting computational efficiency while retaining sample efficiency. 

\begin{figure}
    \centering
    \includegraphics[width=\linewidth]{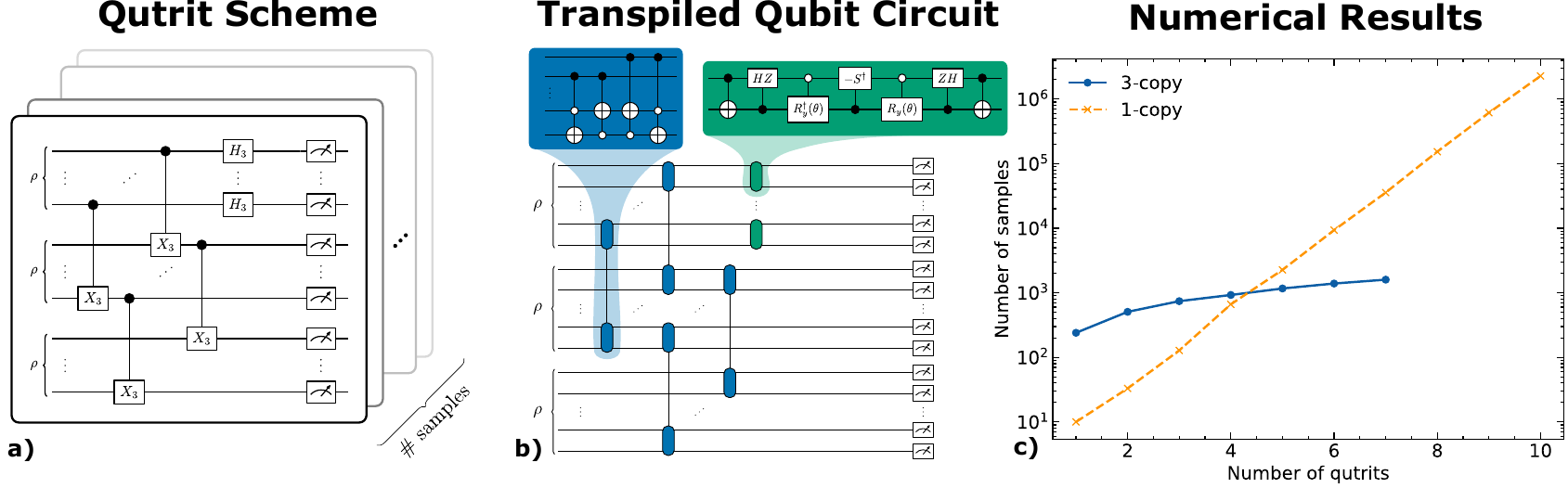}
        \caption{\emph{Learning challenge with degree $c=3$, i.e., an efficient 3-copy strategy exists, but it addresses a degree-2 variant of this learning challenge on the Google Sycamore hard for any 1- and 2-copy strategy.} The underlying learning task is formulated for $n$-qutrit systems ($d=3$) and asks for estimating \emph{all} generalized Pauli observables, also known as Weyl-Heisenberg matrices,  in modulus. \textbf{a)} a qutrit-based 3-copy learning protocol that allows for jointly predicting all third powers of generalized Pauli observables (think: a qutrit extension of the destructive SWAP test for qubit systems). \textbf{b)} a transpilation of this qutrit-based circuit into qubits: $\ket{0}_3 \to \ket{00}$, $\ket{1}_3 \to \ket{01}$, $\ket{2}_3 \to \ket{10}$ (the same can be also done for the underlying learning challenge). \textbf{c)} Numerical simulations that compare the sample complexity of one particular single-copy strategy (orange) -- local classical shadows for qutrit systems~\cite{zhu2025qudit,wilkens2025qudit} -- and the advertised 3-copy strategy (blue) required to complete the learning task with (at least) $70\%$ success probability. The overall plot shape resembles Figure~2 in Ref.~\cite{huang2022quantum} which addressed a degree-2 variant of this learning challenge on the Google Sycamore chip.}
        \label{fig:3c_vs_1c}
\end{figure}

Taken together, these results establish the first infinite hierarchy of multi-copy learning tasks, reveal sharp phase transitions in their sample complexity, and highlight the power of multi-copy quantum information processing as a fundamental resource for quantum learning.

\section{Preliminaries and Notation}\label{sec:preliminaries}

We now turn to the technical development. To set the stage, we first introduce the notation that will be used throughout the paper. With this in place, we revisit the central mathematical object underlying our constructions: the \emph{Weyl–Heisenberg group}, which serves as the natural generalization of the Pauli group to higher dimensions and provides the structural backbone for the learning tasks we study.

\subsection*{Notation}
For an operator $A$ and a multiindex $I=(i_1,\dots,i_m)\in\mathbb{N}^m$ we write $A^{\otimes I}:=A^{i_1}\otimes A^{i_2}\otimes\dots\otimes A^{i_m}$. If $A$ is invertible this notation extends to negative-valued entries of $I$. 
Given a basis $\ket{\psi_1},\dots,\ket{\psi_n}$ of some Hilbert space, for $J\in \{1,\dots,n\}^m$ we write $\ket{\psi_J}:=\ket{\psi_{j_1}}\otimes\ket{\psi_{j_2}}\otimes\dots\otimes\ket{\psi_{j_m}}$. For a multiindex $I=(i_1,\dots,i_m)$ we abbreviate $\abs{I}=\sum_{k=1}^m i_k$ and for the multiindex, which consists of copies of $1$ only, we write $\mathbf{1}=(1,\dots,1)$. Throughout, $d$ always refers to an odd prime, unless indicated differently. All arithmetic involving indices which indicate computational basis elements is then to be understood modulo $d$; we will not always make this explicit. For a natural number $n$ we write $[n]=\{0,1,\dots,n-1\}$. 

\subsection*{The Weyl-Heisenberg operators}
This section introduces the Weyl-Heisenberg group as the natural generalization of the Pauli group to higher dimensions, and briefly reviews its relation to mutually unbiased bases (MUB). Even if the reader is familiar with these, we encourage the reader not to skip this section, as some of our conventions will slightly differ from the standard ones.
The Weyl-Heisenberg operators in dimension $d$ are generated analogously to the qubit Pauli group by a shift operator $X := \sum_{k\in[d]} \kb{k\oplus_d1}{k}$ and a boost operator $Z = \sum_{k\in[d]} = \omega^k \kb{k}{k}$, where $\omega=e^{2\pi\ii/d}$. Here, and in the following, $\oplus_d$ denotes the addition modulo $d$. These elementary Weyl-Heisenberg operators obey the commutation relation $XZ = \omega^{-1}ZX$, as well as $X^d=Z^d=\id$. One can then derive that, up to a suitable factor, all unitary operators generated by $X$ and $Z$ have a projective representative of the form $W(p,q)=X^pZ^q$. It is not hard to verify that all these operators satisfy the following generalized anticommutation relations 
\begin{equation}\label{eq:WH-commutation}
    W(p,q)W(\tilde p,\tilde q)=\omega^{p\tilde q-q\tilde p}W(\tilde p,\tilde q)W(p,q).
\end{equation}
Changing phases in front of operators neither adds complexity in determining expectation values, nor does it affect commutation relations and hence incompatibility of observables. For our purposes, we therefore take the freedom to choose the Weyl-Heisenberg operators in the following way: First, we define 
\begin{equation}
    W_a := XZ^a \text{ for }a\in [d]\quad \text{ and} \quad W_\infty := Z,
\end{equation} 
and let $\WW_a:=\{W_a^k: k \in [d]\}$ be the cyclic group generated by $W_a$. We then obtain a full set of projective representatives by the union $\mathcal{W}^{(d)}:=\bigcup_{a\in[d]}\WW_a\cup\WW_\infty$. We shall omit the superscript when there is no danger of confusion about the underlying dimension. This particular way of grouping the Weyl-Heisenberg operators into distinct cyclic subgroups, which intersect only at the identity, will be convenient for subsequent calculations.  
In particular, to each $\WW_a$ we can associate the orthonormal basis $\mathcal{B}_a=\{\ket{\psi^a_0},\dots,\ket{\psi^a_{d-1}}\}$ of eigenstates, where
\begin{equation}\label{eq:WH-eigenstates}
    \ket{\psi^{a}_j}=\sum_{k\in[d]}\omega^{-jk+a k(k-1)/2}\ket{k}, \text{ for a}\neq\infty,
\end{equation}
and $\mathcal{B}_\infty=\{\ket{k}\}_{k\in [d]}$ is the computational basis. This way we can express each $W_a^k\in\WW_a$ as 
\begin{equation}
    W_a^k = \sum_{j\in[d]}\omega^{kj}\kb{\psi^a_j}{\psi^a_j}.
\end{equation}
In the following, we will often consider the tensor products of local Weyl-Heisenberg operators. In order to distinguish these for the single-qudit Weyl-Heisenberg matrices, we denote them with bold letters $\vec{W}\in(\WW^{(d)})^{\otimes n}=:\WW^{(d),n}$, while we use $W\in\WW^{(d)}$ for the single-site Weyl-Heisenberg matrices for improved distinguishability.

\begin{remark}\label{rem:non-hermitian_obserbvables}
 \emph{Usually, in quantum information, observables are defined as Hermitian operators. Experimentally, we measure observable expectations by performing a measurement in the observable's eigenbasis and weighting the measured probabilities by the corresponding eigenvalues. For Hermitian observables, the eigenvalues of course are always real-valued. However, there is no harm in also considering eigenvalue decompositions with complex-valued eigenvalues, leading to non-Hermitian observables. Expectation values can be formed as long as the operator is diagonalizable.}
 \end{remark}
Following this remark, we see that, in particular, it is justified to treat the Weyl-Heisenberg operators as quantum mechanical observables. Complexity-theoretical results necessarily involve statements about systems that vary in size. One possibility here would be to consider Weyl-Heisenberg tomography for the different underlying qudit systems. However, the structure of the problem changes drastically in this setting, and is not suitable for our purpose. Instead, similarly to \cite{huang2021information,huang2022quantum}, we consider the task of predicting the values of all such Weyl-Heisenberg \emph{strings} on n-qudit systems for fixed $d$. Moreover, to establish a clean separation in the sense of \cref{def:degree}, we consider a slightly simplified task, where we are only interested in the absolute value or \emph{amplitude} of the expected value. Specifically, we consider the following task.
\begin{task}[Weyl-Heisenberg-amplitude-estimation]
    For fixed $\veps>0$, and unknown $n$-partite quantum state $\rho$  with local dimension $d$, approximate $\abs{\tr(\vec W\rho)}$ for all $\vec W\in\WW^{(d),n}$ up to additive error $\veps$. 
\end{task}
In the following, we first derive rigorous lower bounds on the sample complexity for protocols which use fewer than $d$ copies per update step. Our analysis here builds upon the work of \cite{chen2024optimal}. Following up on this, we continue to outline the $d$-copy efficient protocol to solve Weyl-Heisenberg-amplitude estimation in section~\ref{sec:efficient_protocol} and derive the upper bounds, as claimed in \cref{res:main_result}.

\section{\texorpdfstring{Amplitude estimation for $\WW^{(d)}$ is $(d-1)$-copy hard}{Amplitude estimation for W**(d) is (d-1)-copy hard}}\label{sec:lower_bound}

In order to establish lower bounds on sample complexity, one must show that a task remains hard for any conceivable protocol, including those with adaptive strategies where each measurement can be chosen in response to the entire prior measurement history. A powerful abstraction for reasoning about such protocols is the framework of \emph{learning trees}.

The idea is as follows. We model the learning algorithm as a machine equipped with an internal memory and the ability to perform measurements on quantum states. At time step $t$, given a memory state $u$, the machine performs a measurement $M_u$ on (copies of) the unknown quantum state $\rho$. Depending on the outcome $a \in \{a_1, \dots, a_{m(u)}\}$, the machine updates its memory to a new node $v_a$. This update is represented by a directed edge from $u$ to $v_a$.  

Because we are interested in lower bounds on sample complexity, we may, without loss of generality, assume the machine has unbounded memory. Consequently, no two directed edges need to merge into the same node, since such merging would mean the algorithm discards information. The underlying structure is therefore a tree, not just a directed graph.  

The problem of bounding the depth of such learning trees required to solve a given task with high success probability has been extensively studied~\cite{chen2024optimal,huang2022quantum}. However, the learning tree framework is most naturally suited to \emph{hypothesis testing}. To exploit this approach, we must therefore identify a hypothesis testing problem that reduces to Weyl-Heisenberg amplitude estimation while still yielding stringent lower bounds. Following the approach of~\cite{huang2022quantum,chen2024optimal}, a particularly effective choice is that of \emph{many-versus-one distinguishing tasks}.  
These tasks are hypothesis tests in which one must decide between a single null hypothesis and a collection of alternative hypotheses, hence the name \emph{many-versus-one}. In the quantum setting, such problems have been analyzed in great detail in~Ref. \cite{chen2024optimal}, leading to general lower bounds on their sample complexity. For our purposes, we introduce the instance most relevant to our construction directly.  

\begin{task}[Many-versus-one distinguishing task] \label{task:mvo_separation}
Fix $0<\veps\leq1/6$. Distinguish between the following two hypotheses, each occurring with prior probability $1/2$:  
\begin{itemize}
    \item \textit{Null hypothesis:} the state is maximally mixed, $\rho = \rho_m:=\id/d^n$.
    \item \textit{Alternative hypothesis:} the state is a fixed but random element of the ensemble
    \begin{equation}\label{eq:Mvs1_ensemble}
        \Bigl\{\, \rho^\veps_\vec{W} := \tfrac{1}{d^n}\bigl(\id + 3\veps(\vec W+\vec W^\dagger)\bigr) \,\Bigr\}_{\id\neq W\in\WW^{(d),n}}.
    \end{equation}
\end{itemize}   
\end{task}

The constraint $\veps \leq 1/6$ ensures that $\rho^\veps_\vec{W}$ is always positive semidefinite, but can be relaxed for certain concrete dimensions, e.g., to $\veps \leq 1/3$ when $d=3$.  
The following theorem is a direct corollary of Theorem~1 in~\cite{chen2024optimal} applied to our setting. It relates the sample complexity of solving the many-versus-one-distinguishing \cref{task:mvo_separation} to the following quantity  
\begin{equation}\label{eq:delta_quantity}
\delta_{\mathcal{\veps},c}:=\min_{\mu\in\mathscr{P}(\WW^{(d),n})}\max_{\mathcal{M}}\EE_{\vec{W}\sim\mu}[\chi^2_\mathcal{M}((\rho_\vec{W}^\veps)^{\otimes c}||\rho_m^{\otimes c})].
   \end{equation}
Here, $c$ refers to the number of state copies that can be processed simultaneously, the maximization runs over all (finite) POVMs $\mathcal{M}$, and the minimization runs over all probability distributions $\mu$ on $\WW^{(d),n}$. $\chi^2_\mathcal{M}$ refers to the classical $\chi^2$-divergence between the probability distributions obtained from the respective states via the measurement $\mathcal{M}$. Recall that for discrete probability distributions, the $\chi^2$-divergence is defined as 
\[\chi^2(p||q):=\sum_j \left(\frac{p_j-q_j}{q_j}\right)^2q_j.\]
With these clarifications, we are now ready to state the Theorem. 

\begin{theorem}[Theorem 1 in \cite{chen2024optimal}]\label{thm_Master}
   Consider the many-versus-one distinguishing task between the maximally mixed state $\rho_m$, and the ensemble of states $\{\rho^\veps_W\}$ as in \cref{eq:Mvs1_ensemble}. Any learning protocol which can process up to $c$ copies of the distributed state $\rho$ simultaneously and solve said task correctly with probability $p>1/2$, needs at least $\Omega(c/\delta_{\veps,c})$ copies of $\rho$, with $\delta_{\veps,c}$ as specified in \cref{eq:delta_quantity}
\end{theorem}

In order to apply these results to our problem of lower bounding the sample complexity for  Weyl-Heisenberg amplitude estimation, we need to reduce the former to the latter. While the intuition should be relatively straightforward, we nevertheless give a short proof to convince ourselves of the correct choice of parameters in the following lemma.
\begin{lemma}[Reduction to separation instance]\label{lemma:reduction}
    Assume that we can solve the Weyl-Heisenberg amplitude estimation with high probability for the precision parameter $\frac{1}{6}\geq \veps>0$. Then, with the same probability, we can also solve the separation instance of the many-versus-one learning task specified in \cref{task:mvo_separation}. In particular, the sample complexity for solving WH-amplitude estimation is lower bounded by $\mathcal{O}(c/\delta_{c,\veps})$ with $\delta_{\veps,c}$ as in \cref{eq:delta_quantity}.
\end{lemma}
\begin{proof}
    If one is given estimates $u_\vec{W}$ such that $\max_\vec{W} \abs{u_\vec{W}-\abs{\tr(\vec{W}\rho)}}\leq\veps$, it is easy to see that we can distinguish the two cases displayed in Task~\ref{task:mvo_separation}. 
    Case 1 ($\rho = \id/d^n)$ implies $\mathrm{tr}(\vec{W} \rho)=0$ for all $\vec{W}$ and therefore $\abs{u_\vec{W}-\tr[\rho_m\vec{W}]}=\abs{u_\vec{W}}\leq\veps$.
    Case 2 instead yields substantially larger values for exactly two Weyl-Heisenberg operators ($\vec{W}$ and $\vec{W}^\dagger$): $\tr[\vec{W}\rho_\vec{W}]=\tr[\vec{W}^\dagger\rho_\vec{W}]=3\veps$ and we find that $\abs{u_\vec{W}}\geq\abs{\tr[\vec{W}\rho]}-\tr[\vec{W}\rho]\geq 2\veps$. Note that when $d$ is even, we might end up with some cases where $\vec{W}=\vec{W}^\dagger$ and we have $ \tr[\vec{W}^\dagger\rho_\vec{W}]=6\veps$ in these cases. However, this does not fundamentally change our argument, as it simply leads to a larger separation.
    Therefore, if $\max_\vec{W} u_\vec{W}\leq\veps$ we opt for the null hypothesis, and for the alternative hypothesis otherwise. 
    In particular, the probability of solving the hypothesis test correctly is equal to the probability of obtaining $\veps$-correct estimates $u_\vec{W}$ for all $\abs{\tr[\vec{W}\rho]}$ in the considered instance.
\end{proof}

This reduces the problem of deriving sample complexity lower bounds to upper bounding the quantity $\delta_{c,\veps}$ from Eq.~\eqref{eq:delta_quantity}. Inserting the $c$-fold copies of the two state families yields
    \begin{equation}
\delta_{c,\veps}=\min_{\mu\in\mathscr{P}(\WW^{(d),n})}\max_\mathcal{M}\EE_{\vec{W}\sim\mu}\left[\chi^2_M\left(\left(\frac{I+3\veps (\vec{W}+\vec{W}^\dagger)}{d^n}\right)^{\otimes c}\abs{\vphantom{\sum_{j=0}}}\left(\frac{I}{d^n}\right)^{\otimes c}\right)\right].
\end{equation}
To bound this expression from above, we proceed in essentially two steps. First, we show that by picking $\mu=\nu^{\otimes n}$ for any probability distribution $\nu$ over $\WW^{(d)}$, we can upper bound $\delta_{c,\veps}$  as follows:
    \begin{equation}\label{eq:auxiliary_bound_1}
        \delta_{c,\veps}\leq\min_{\nu\in\mathscr{P}(\WW^{(d)})}\max_{1\leq m\leq c}\max_{\tau\in\{\pm 1\}^{2m}}\norm{\vphantom{\sum}\EE_{W\sim\nu}[W^{\otimes\tau}]}^n(6c\veps \exp(6c\veps))^2.
    \end{equation}
The derivation of this bound, which can be found in \cref{lemma:first_delta_bound}, is very technical and in large parts analogous to a calculation in \cite{chen2024optimal}. A subtle difference compared to the scenario of \cite{chen2024optimal} is, that since the state $\rho^\veps_W$ also involves the inverse $W^\dagger$ of $W$ to account for the non-hermiticity of $W$, we need to find bounds for all the averages $M_\tau = \sum_{W\in\hat{\WW}}W^{\otimes\tau}$ for the various $\tau\in\{\pm1\}^{2m}$, not only for $W\mapsto W^{\otimes 2m}$. The second step of the proof involves estimating the operator norm of the averages for some suitable choice of distribution $\nu$. In \cref{sec:proofs} we derive a series of intermediate results \cref{prop:prime_power_result}, \cref{lemma:qubit_Pauli_twirling}, ultimately contributing to the following norm bounds, from which the sample complexity can be derived by virtue of \cref{thm_Master}.
\begin{proposition}\label{prop:composite_norm_bounds}
    Let $d\geq 2$ be an integer with prime factorization $d=p_1^{r_1}\dots p_l^{r_l}$ in increasing order $p_1<\dots<p_l$. Then there exists a probability distribution $\nu$ over $\WW^{(d)}$ such that  for any $m<d$ and $\tau\in\{\pm1\}^{2m}$ we have
    \begin{equation}
        \norm{\EE_{W\sim\nu}[W^{\otimes\tau}]}\leq\max_j \chi(p_j^{r_j}),\text{ where }\chi(p^{r}):=\begin{cases}
            \frac{1}{2},&p=2\text{ and }r=1\\
            \frac{1}{p-1}, &p\neq 2\text{ and }r=1\\
            \frac{2p^{r-1}}{p^r-1}, &\text{ else}.
        \end{cases}
    \end{equation}
    In particular, if $d$ is not divisible by four, then we find in any case $\norm{\vphantom{\sum_j}\EE_{W\sim\nu}[W^{\otimes\tau}]}\leq\frac{3}{4}<1$.
\end{proposition}
We prove this claim in \cref{sec:proofs}, after \cref{prop:prime_power_result}. \noindent Altogether, we have proven the following theorem, which constitutes the second part of \cref{res:main_result}.
\begin{theorem}
    Let $d\geq 2$ be an integer not divisible by four, $c<d$ and $\veps<\frac{1}{6}$. Any $c$-copy protocol which solves the Weyl-Heisenberg amplitude estimation task for an arbitrary $n$-qudit state $\rho$ and hence, by extension, the many-versus-one distinguishing  \cref{task:mvo_separation}, requires a number of samples of $\rho$ growing at least as 
    \begin{equation}
        N\geq \Omega\left(\left(\frac{4}{3}\right)^nc^{-1}\veps^{-2}\exp(-c\veps)\right).
        \end{equation} 
\end{theorem}
The bound resulting from the choice of $b=\frac{4}{3}$ as the exponential base in the theorem above is generally very loose. First of all, as evident from \cref{prop:composite_norm_bounds}, it can be improved for each $d$ individually. In particular, let us note the following special cases: When $d$ is prime, we can choose $b=d-1$. When $d$ is square-free and odd, we have $b=p-1$, with $p$ being the smallest prime factor of $d$; and $b=2$ for $d$ square-free and even. Moreover, we can in principle obtain a more fine-grained dependence on $c$, as evident from \cref{prop:composite_norm_bounds}, by analyzing which prime (powers) can appear as a factor for numbers $m\leq c$.
However, properly keeping track of all the factors as a function of $c$ becomes rather cumbersome, and we leave it to the remark that the stated bounds are in general suboptimal. 
On the other hand, we note that for higher powers of two, the lower bound in \cref{prop:composite_norm_bounds} becomes trivial. In particular, for all investigated measures $\nu$ on $\WW^{(d)}$ for $d=2^r,r>1$ which exhibit a high degree of symmetry, and therefore - intuitively - should make the problem hard, we already found that $\norm{\EE_{W\sim\nu}[W^{\otimes (d/2)}]}=1$, thus breaking the exponential scaling. However, so far it is unclear whether the task actually becomes easy already for $d/2$-copy measurements in these instances, potentially reflecting the fact that the Clifford and Weyl-Heisenberg groups behave quite differently when $d$ is odd \cite{Gross2006hudso}, or if there is in fact an exponential separation and our analysis is simply insufficient to reveal it, which we leave open for further research.

\section{\texorpdfstring{Amplitude estimation for $\WW^{(d),n}$ is $d$-copy efficient}{Amplitude estimation for W**(d) is d-copy efficient}}\label{sec:efficient_protocol}

Similar to the Pauli-tomography task, one can bypass the exponential scaling of the sample complexity by a sufficiently large quantum memory that permits multi-copy measurements. In the same way, as two copies are sufficient for Pauli tomography, performing entangling measurements across $d$ copies of the state allows one to measure all Weyl-Heisenberg strings simultaneously. However, as with Pauli tomography, this partially blurs the information and only allows one to predict the absolute value, i.e., the amplitude, reliably. If one additionally wishes to extract information about the phase, one has to go through a somewhat involved post-processing scheme, similar to the one proposed in \cite{king2025triply} which is described in \cref{sec:extension_to_full_tomography}. 
One might be tempted to simply estimate $\abs{\tr[\vec{W}\rho]}$ via the conjugate representation $\vec{W}\otimes\bar{\vec{W}}$. However, this would require access to $\bar{\rho}$ as an additional resource, which cannot be obtained from $\rho$ via any physical operation. The power of having access to $\bar\rho$ as an additional resource is discussed extensively in \cite{king2024exponential}.
If, however, the unknown state is assumed or believed to be real-valued, then $\bar{\rho}=\rho$, and this observation provides a significant shortcut. On the other hand, complex numbers are an intrinsic feature of quantum mechanics and generic states are certainly not exclusively real-valued~\cite{renou2021quantum}, and this observation does not contradict our overall conclusion.

Let us first consider the case where $n=1$, i.e.\ a single $d$-dimensional qudit. We observe that according to \cref{eq:WH-commutation}, any two elements of $\WW$ commute up to a power of $\omega$ as a prefactor. Hence, under the $d$-th power tensor representation $\Phi^d: \vec{W}\mapsto \vec{W}^{\otimes d}$, all elements in $\WW$ commute. In particular, there is an \ac{ONB}, in which all $\vec{W}^{\otimes d},\,\vec{W}\in\WW$ are simultaneously diagonal. A relatively straightforward calculation shows that this diagonalization is achieved by the following natural generalization of the Bell basis to $d$-qudit systems
\[\ket{\psi_{I,q}}=\frac{1}{\sqrt{d }}\sum_{k=0}^{d-1}\omega^{kq}\ket{(0,I)+ k\mathbf{1}}, \text{ where }q\in[d],I\in[d]^{d-1}.\]
It is also easy to check that the states $\ket{\psi_{I,q}}$ for $I\in[d]^{d-1},\,q\in[d]\}$ form an \ac{ONB} of $\left(\CC^d\right)^{\otimes d} $. What is more, these orthonormal states are also eigenstates of all $\vec{W}^{\otimes d}$. For this it is sufficient to consider the generators $X,Z\in\WW$:
\begin{align}
    \label{eq:X-eigenvalues}X^{\otimes d}\ket{\psi_{I,q}} =& \frac{1}{\sqrt{d}}\sum_{k\in[d]} \omega^{kq}\ket{(0,I)+ k\mathbf{1}}=\omega^{-q}\ket{\psi_{I,q}},\\
    \label{eq:Z-eigenvalues}Z^{\otimes d}\ket{\psi_{I,q}} =& \frac{1}{\sqrt{d}} \sum_{k\in[d]}\omega^{kq}\omega^{\abs{I}+dk}\ket{(0,I)+k\mathbf{1}}=\omega^{|I|}\ket{\psi_{I,q}}.
\end{align}
Now observe the following: The dependence of the eigenvalue of any $(X^aZ^b)^{\otimes d}$ on the basis state $\ket{\psi_{I,q}}$ can be significantly coarse-grained. Specifically, the information on the multiindex $I$ gets blurred up to the value $\abs{I}$. Therefore, in order to determine the outcomes of all Weyl-Heisenberg observables, it suffices to consider the following projective $d^2$-outcome measurement 
\begin{equation}
  \{\Pi_{s,q}|s,q\in[d]\}\text{ where } \Pi_{s,q}=\sum_{I\in[d]^{d-1},\abs{I}=_ds}\kb{\psi_{I,q}}{\psi_{I,q}},  
\end{equation}
rather than the full generalized Bell-basis measurement. This measurement primitive readily extends from the single-qudit case to the $ n$-qudit case by performing the (coarse-grained) Bell-basis measurements in parallel on the respective $d$-qudit clusters. Specifically, for $\mathbf{q}\in[d]^n, \mathbf{s}\in[d]^n$ we consider the PVM $\{\Pi_{\mathbf{q},\mathbf{s}}\}$ with $\Pi_{\mathbf{q},\mathbf{s}}:=\bigotimes_{k=1}^n \Pi_{q_k,s_k}.$
In particular, the expectations of the WH-string are determined as
\[\tr[(X^{\otimes\mathbf{a}}Z^{\otimes\mathbf{b}})^{\otimes d}\rho^{\otimes d}]=\sum_{\mathbf{q},\mathbf{s}}\omega^{\langle\mathbf{b},\mathbf{s}\rangle-\langle\mathbf{a},\mathbf{q}\rangle}\,\tr(\Pi_{\mathbf{q},\mathbf{s}}\rho^{\otimes d}).\]
From (independent repetitions of) this single basis measurement on $dn$ qudits in total, the values of all observables $\mathrm{tr} \left( \vec{W}^{\otimes d} \rho^{\otimes d}\right) = \mathrm{tr} \left( \vec{W} \rho\right)^d$ can be jointly estimated. This, in turn, allows us to predict the absolute values $\abs{\tr[\vec{W}\rho]}$ up to additive error $\veps$, by estimating $\tr[\vec{W}^{\otimes d}\rho^{\otimes d}]$ up to precision $\veps^d$ for all $\vec{W}\in\WW^n$.
Note that, aside from the ability to process $d$, $n$-qudit states, the protocol does not require quantum resources that hinder practical applicability. Specifically, the circuits performing the generalized Bell-state measurements can be executed in parallel as schematically depicted in \cref{fig:3c_vs_1c}a), and can be implemented in a way using at most two-qudit gates as non-local gates. In particular, the depth remains unaffected as a function of $n$. Explicit quantum circuits for the execution of these measurements, as well as a transpilation procedure for embedding them into qubit systems for the case of qu\emph{trits}, are depicted in \cref{fig:3c_vs_1c}b).

\subsection*{Sample complexity}
For $\vec{W} \in\mathcal{W}^{(d),n}$ let $X^{(i)}_\vec{W}$ denote the associated random variable taking values in $\{1,\omega,\dots,\omega^{d-1}\}$ according to the measured outcome in round $i$. In order to obtain estimates on the sample complexity, we employ Hoeffding's inequality (see \cref{lemma:Hoeffding}). For this, we first observe that if a complex number $z$ satisfies $\abs{z}\geq \gamma$ for some $\gamma>0$, we have $\max\{\abs{\Re(z)},\abs{\Im(z)}\}\geq\frac{\gamma}{\sqrt{2}}$. By employing a union bound for maximizing over real and imaginary parts as well as all Weyl-Heisenberg matrices, we can bound the deviation from the actual mean after $N$ trials as follows 
\begin{align*}
  &\Pr\left[\max_{W\in\WW^{(d)}}\abs{\sum_{i=1}^N \frac{X_\vec{W}^{(i)} -\tr(\vec{W}\rho)^d}{N}}\geq \veps\right]\leq  \sum_{\vec{W}\in\WW^{(d)}}\Pr\left[\max_{F\in\Re,\Im}\abs{\sum_{i=1}^N \frac{F[X_i -\tr(\vec{W}\rho)^d]}{N}}\geq \frac{\veps}{\sqrt{2}}\right]\\
  \leq& \sum_{\vec{W}\in\WW^{(d)}}\sum_{F=\Re,\Im}\Pr\left[\abs{\sum_{i=1}^N \frac{F[X_i -\tr(\vec{W}\rho)^d]}{N}}\geq \frac{\veps}{\sqrt{2}}\right] \leq  2d^{2n}2e^{-\veps^2N/4},
\end{align*}
where we used that $\abs{X_i-\mathbb{E}[X_i]}=\abs{X_i-\tr[\rho \vec{W}]^d}\leq 2$ in the last step. Since $\abs{\abs{\tr(\vec{W}\rho)}-r}\leq\abs{\tr(\vec{W}\rho)^d-r^d}^{1/d}$ for any $0\leq r\leq 1$, this shows that altogether $S=dN\geq \mathcal{O}(  d\veps^{-2d}(n\log{d}+\log(\delta^{-1}))$ state copies are sufficient to determine all $\abs{\tr[\vec{W}\rho]}$ up to $\veps$-precision simultaneously with confidence $1-\delta$. 
\subsection*{Quantum circuit complexity}
As we showed in the previous section, it is possible to estimate all Weyl-Heisenberg-amplitudes \emph{sample-efficiently}. Overall, however, precious little would be gained if this advantage was only achieved by significantly increasing the resource requirements on the quantum hardware. As we clarified already, we perform measurements across a \emph{constant} number of states, which is a significant advantage over shadow tomography. Additionally, we observe that while the measurement must act as an entangling measurement on the global system of dimension $d^{dn}$, the circuit size scales only linearly in $n$, since the diagonal basis can be prepared as the $n$-fold tensor product of generalized Bell states on every group of $d$ qudits. Moreover, preparing the generalized Bell states (as in \cref{fig:3c_vs_1c}a) in terms of single- and two-qudit Cliffords is straightforward and requires circuits of size $\mathcal{O}(dn)$ and depth $\mathcal{O}(d)$. 
While the depth and size of the circuits preparing individual generalized Bell-basis measurements will in general change depending on the underlying $d$ if one were interested in transpiling the qudit architecture to qubit systems, this does not affect the linear scaling in $n$. To summarize our findings in this section so far, we have also proven the first part of \cref{res:main_result}. 

\begin{theorem}
By performing parallel generalized Bell measurements, we find a $d$-copy efficient protocol, which uses $\mathcal{O}(nd\log(d)\veps^{-2d})$ samples of $\rho$ to solve the Weyl-Heisenberg amplitude estimation up to additive error $\veps>0$. The quantum circuits to realize these measurements in terms of single- and two-qudit Clifford gates have size $\mathcal{O}(dn)$ and depth $\mathcal{O}(d)$, see \cref{sec:q_circuits}.
\end{theorem}

\section{Conclusion and Outlook}
We have successfully demonstrated the existence of degree-$d$ learning tasks for every prime number $d$, that is, learning tasks which are hard to solve if the measurement device can process only up to $d-1$ copies of an unknown state at one time, but become efficiently solvable, reducing to a linear scaling in the system size, if $d$-copies can be measured together.\\

We achieved this by the natural generalization of the degree-2 task of determining the amplitudes of all Pauli strings, as discussed in \cite{huang2022quantum,chen2024optimal}, to the Weyl-Heisenberg group. The generalization of the efficient $d$-copy scheme is pretty much straightforward by considering the natural extension of Bell-basis measurements to qudit systems, as we demonstrated in \cref{sec:efficient_protocol}. On the other hand, we reduced the learning task to a hypothesis test in the framework of many-versus-one-distinguishing tasks, and exploited the machinery developed in \cite{chen2024optimal} to derive lower bounds on the number of iterations/measurements necessary to solve this type of hypothesis test reliably. 

Finally, we also briefly outlined how one can extend this scheme to the full tomography of all Weyl-Heisenberg strings, at the expense of exponentially increasing computational cost, while remaining sample efficient. 

Notably, our analysis remains inconclusive for the degree of estimating the Weyl-Heisenberg amplitudes when it involves higher powers of two. We leave it open for further research whether this is due to the structure of the Weyl-Heisenberg group in these dimensions, or if there is an actual separation, which we are simply unable to uncover with our methods.

Our learning challenges can either be formulated in terms of Weyl-Heisenberg shadow tomography on $n$-partite qudit systems ($d \geq 2$), or they can readily be transpiled into equivalent learning challenges on $O(\log(d)n)$-qubit systems. The latter is conceptually closer to the thematically related, but independent, submission~\cite{ye2025exponential}, as well as previous work~\cite{huang2021information,huang2022quantum} (which we recover in the special case of $d=2$). The former interpretation, however, opens up interesting future research directions as well. Qudit-based classical simulation techniques could be used to analyze the susceptibility of these learning strategies vis-à-vis noise. This can, in turn, inform the design of qudit-based noise mitigation strategies that may then allow an actual implementation of Weyl-Heisenberg tomography on a native qudit-based quantum processor~\cite{ringbauer2022qudit}. It would also be interesting to compare the performance of such genuine qudit-based implementations with an equivalent learning protocol transpiled for qubit-based architectures. 

On a more conceptual level, we believe that our generalizations of the two-copy learning results from Refs.~\cite{huang2021information,huang2022quantum} can also form the basis of a generalization of several more sophisticated learning protocols that build upon two-copy Bell sampling (also known as the destructive SWAP test). Concrete examples are triply-efficient shadow tomography~\cite{king2025triply} (which we have already partially addressed in Algorithm~\ref{alg:protocol} above), as well as alternative protocols for auxiliary-free replica shadow estimation~\cite{liu2024auxiliary}.    Another observation that opens up possibilities for further research is that the efficient protocol in \cref{sec:efficient_protocol} does not actually perform any genuine $d$-partite entangling operations. In fact, the entire protocol can be cast as a quantum streaming algorithm~\cite{kallaugher2022quantum,Kallaugher2025how}, where states are sequentially distributed, and a two-state quantum instrument is applied, destroying one copy, while the remaining part is stored in a quantum memory and processed with the next copy. Under this viewpoint, our work also identifies a new family of problems for which streaming-type algorithms exhibit a provably exponential advantage, where the quantum memory can be erased after every few steps, thereby putting only mild requirements on the necessary coherence times.

\section{Acknowledgements}
We thank Antonio and Francesco Anna Mele for discussions. This research was funded by the Deutsche Forschungsgemeinschaft
(DFG, German Research Foundation), project numbers 441423094, 236615297 - SFB 1119, the Austrian Science Fund (FWF) via the SFB BeyondC (10.55776/FG7) and an FWF START award, as well as the European Research Council (ERC) via the Starting grant q-shadows.

\appendix
\newpage
\section*{Appendix}
The Appendix is organized as follows: \Cref{sec:qudit_Clifford} briefly reviews the qudit-Clifford group, supplemental to the introduction of the Weyl-Heisenberg group in the main text. In \cref{sec:proofs} we provide the proofs of some results which we claimed but did not prove in the main part, where other technical results are outsourced further to \cref{sec:lemmata}. Finally, \cref{sec:q_circuits} features the quantum circuits necessary to run the $d$-copy-efficient protocol from \cref{sec:efficient_protocol}.

\section{Qudit quantum computing}\label{sec:qudit_Clifford}
Quantum computing in qubits makes heavy use of the Pauli matrices, which are "classical" quantum operations, in the sense that if one splits a qubit into amplitude and phase information, they merely act as flip operators; they themselves can neither create nor destroy coherence in the computational basis. In the main text, we already provided the analog of the Pauli group in higher dimensions, the Weyl-Heisenberg group. Similarly to the qubit case, we can then proceed to define the qudit-Clifford group as the symmetry group of the Weyl-Heisenberg matrices.
\begin{equation}
    \mathcal{C}l=\{U|UW U^\dagger \in\WW\text{ for all }W\in\WW\}.
\end{equation}
Similar to the qubit case, the Clifford group is a maximal finite subgroup of the unitary group $\mathrm{U}(d)$. In particular, in the single-qudit case it has only two generators, which resemble the Hadamard $H$ and the $S$-gate. For the $n$-qudit Clifford group, next to the local generators $S$ and $H$ on each subsystem, one additionally needs $n-1$ generators which generalize the role of the controlled-NOT operation in the multi-qudit case; see, e.g., \cite{jafarzadeh2020randomized}. We now explicitly state these generators and some of their properties inside the Clifford group.
First, consider the Fourier gate, as the analog of the Hadamard gate, defined via
\begin{equation}\label{eq:H-gate_Def}
    F = \frac{1}{\sqrt{d}}\sum_{k,j\in[d]}\omega^{kj}\kb{k}{j}.
\end{equation}
It has order four $F^4=\mathds{1}$, and interchanges between the $X$ and $Z$ bases such that $FZF^\dagger =X$. The analogue to the $S$-gate in the qubit Clifford group, is the following \emph{quadratic} phase gate
\begin{equation}\label{eq:S-gate_Def}
    Q=\sum_{j\in[d]} \omega^{j(j-1)/2}\kb{j}{j}.
\end{equation}
It satisfies $Q^d=\mathds{1}$, and together with $H$ generates $Z$ via $Z=Q^{d-1}F^2QF^2$. In particular, the operator $S$ is constructed such that $\ket{\psi^a_j}=S^a\ket{\psi^0_j}$ for $a\neq\infty$.
Additionally, we have the following two-qudit gate, which serves as a generalization of the CNOT gate, where the addition on the second qudit is now simply performed modulo $d$:
\begin{equation}\label{eq:CX-gate_Def}
    CX = \sum_{k,l\in[d]}\kb{k}{k}\otimes\kb{k+l}{l}.
\end{equation}
Together, the local generators $Q$ and $F$ on every qudit and the two-qudit gates $CX^{(j,j+1)}$ for $j=1,\dots,n-1$ generate the entire $n$-qudit Clifford group.

\section{Sample-efficient extension to full WH-tomography}\label{sec:extension_to_full_tomography}
The protocol outlined in \cref{sec:efficient_protocol} only provides estimates of the amplitude of $\tr(\vec W\rho)$ for all Weyl-Heisenberg strings $\vec W$, rather than the actual value. If, for whatever application one is interested in, this is not enough and one additionally needs the phase information, there is a way to expand the protocol to estimate the actual value of all $\tr(\vec W\rho)$ in a sample-efficient manner. However, this comes at the expense of requiring access to significantly increased classical and quantum computational resources. 
The idea is to prepare copies of a known state $\sigma$, which has the property that $\abs{\tr(\vec W\sigma)}$ is significant whenever $\abs{\tr(\vec W\rho)}$ is, as determined beforehand via WH-amplitude estimation. Specifically, given $d-1$ copies of such a \emph{mimicking state} $\sigma$ together with one copy of $\rho$, we can jointly estimate $\tr(\vec W^{\otimes d}\sigma^{\otimes d-1}\otimes\rho)$ for all $\vec W$. Then, since the values of $\tr(\vec W\sigma)$ are known, we can efficiently determine the actual values of all $\tr(\vec W\rho)$. The computational bottleneck for this procedure is clearly to find such a mimicking state $\sigma$. This can be achieved based on Hamiltonian update rules via the matrix multiplicative weight algorithm, following ideas of \cite{king2025triply,tran2025one}. Note that our scenario again requires only minor adaptations compared to these mentioned results. The modified algorithm, the correctness of which we prove formally in \cref{lemma:correctness}, is stated below as \cref{alg:protocol}.
\begin{algorithm}[hbt]
\caption{Mimicking state construction}\label{alg:protocol}
\flushleft{\textbf{Given:} estimates $u_\vec{W}$ s.t. $\abs{u_\vec{W}-\abs{\tr(\rho \vec{W})}}\leq\veps/6$.\\
\textbf{Output:} $\sigma$, such that $\abs{\tr(\vec{W}\sigma)}\geq \veps/3$ whenever $u_\vec{W}\geq\veps.$\\

\textbf{Initialise:} $T=\lfloor 75 n/\veps^2\rfloor+2$; $\;\beta = \sqrt{n/T}$; $\;\sigma_0=\mathds{1}/d^n$.\\
}
For $t=0,\dots,T-1$ do the following:
\begin{enumerate}
    \item Search for $\vec{W}_t\in \WW^n$ s.t. $u_t:=u_{\vec{W}_t}\geq \veps$ and $\abs{\abs{\tr(\vec{W}_t \sigma_t)}-u_t}\geq2\veps/3$.
    \item If no such $\vec{W}_t$ is found, output $\sigma=\sigma_t$.
    \item Otherwise: 
    \begin{enumerate}
    \item Use $\mathcal{O}(\log(T)/\veps^2)$ copies of $\rho$ to produce an estimate $z_t$ such that $\abs{\tr(\vec{W}_t\rho)-z_t}\leq\veps/25$ with probability $p\geq 1-0.01/\abs{T}$.
    \item Determine $\argmax_{F\in\{\Re,\Im\}}\abs{F(\tr(\vec{W}_t\sigma_t)-z_t)}$.\\
            If $\argmax=\Re$ then: $\hat{\vec{W}}_t:=\frac{\vec{W}_t+\vec{W}_t^\dagger}{2}$, and $\hat{z}_t:=\Re[z_t]$\\
            If $\argmax=\Im$ then: $\hat{\vec{W}}_t:=\frac{\vec{W}_t-\vec{W}_t^\dagger}{2i}$, and $\hat{z}_t:=\Im[z_t]$\\
            $\Gamma_t:=\sgn(\tr(\hat{\vec{W}}_t\sigma_t)-\hat{z}_t)\cdot\hat{\vec{W}}_t$
    \item $\sigma_{t+1}:=\frac{1}{\mathcal{N}}\exp\left(-\beta\sum_{\tau=1}^t \Gamma_\tau\right)$ with suitable normalization factor $\mathcal{N}$.
    \end{enumerate}
    \end{enumerate}
  Output $\sigma=\sigma_T$.
\end{algorithm}
Finally, let us briefly analyze how the various error thresholds propagate during the procedure.
As we prove in the Appendix, given $\veps>0$ and $\veps/6$-accurate estimates of all $\abs{\tr(\vec{W}\rho)}$, with high probability, \cref{alg:protocol} outputs a state $\sigma$, such that $\abs{\tr(\vec{W}\rho)}\geq\veps\Rightarrow \abs{\tr(\vec{W}\sigma)}\geq\veps/3.$ Let $z_\vec{W}$ be $\veps^d$-accurate estimators  of $\tr(\vec{W}^{\otimes d} \sigma^{d-1}\otimes\rho)$. Then

\[\abs{\tr[\vec{W}\rho]-\frac{z}{\tr[\vec{W}\sigma]^{d-1}}}=\frac{\abs{\tr[\vec{W}\sigma]^{d-1}\tr[\vec{W}\rho]-z}}{\abs{\tr[\vec{W}\sigma]}^{d-1}}\leq\frac{\veps^{d}}{(1/3)^{d-1}\veps^{d-1}}=3^{d-1}\veps.\]
In particular, we see that the procedure outlined in this section requires only $\mathcal{O}(nd\log(d)\veps^{-2d})$ samples of $\rho$ over all involved steps to estimate all the values $\tr(\vec{W}\rho)$ up to additive error $\veps>0$.

\section{Outsourced Proofs from Main part}\label{sec:proofs}
We begin by providing the proofs of the two main steps for deriving the lower bound on the sample complexity in \cref{sec:lower_bound}.
\begin{techlemma}[Compare with \cite{chen2024optimal}, Thm. 15, slightly extended and modified version thereof]\label{lemma:first_delta_bound}\hfill
We can bound $\delta_{c,\veps}$ as defined in \cref{eq:delta_quantity} by
    \begin{equation}
        \delta_{c,\veps}
        \leq \min_{\nu\in\mathscr{P}(\WW^{(d)})}\max_{1\leq c\leq m}\max_{\tau\in\{\pm 1\}^{2m}}\norm{\EE_{W\sim\nu}[W^{\otimes\tau}]}_\infty^n (6c\veps e^{6c\veps})^2,
    \end{equation}
    where the minimization runs over all $\nu\in\mathscr{P}(\WW^{(d)})$, the set of all probability measures on $\WW^{(d)}$.
    \end{techlemma}
 \begin{proof}
    The proof is essentially analogous to the one of Theorem 15 in \cite{chen2024optimal}. Due to subtle differences in several places, however, we conduct the proof here nonetheless. First of all, we spell out the $\chi^2$-divergence for fixed $\vec W$ and measurement $M=\{M_a\}_{a\in A}$ acting on $\left(\CC^{dn}\right)^c$ as follows:
    \begin{align*}
        \chi^2_M\left((\rho_\vec{W}^{\veps})^{\otimes c}\left|\right|\rho_m^{\otimes c}\right)&=\sum_{a\in A}\tr(M_a \rho_m^{\otimes c})\left(\frac{\tr(M_a (\rho_\vec{W}^\veps)^{\otimes c})}{\tr(M_a\rho_m^{\otimes c})}-1\right)^2\\
        &=\sum_{a\in A}\frac{\tr(M_a)}{d^{cn}}\left(\frac{\tr(M_a (\openone+3\veps(\vec W+\vec W^\dagger))^{\otimes c})}{\tr(M_a)}-1\right)^2\\
        &=\sum_{a\in A}\frac{1}{\tr(M_a) d^{cn}}\left(\sum_{0\neq S\in[2]^c}(3\veps)^{\abs{S}}\tr(M_a (\vec W+\vec W^\dagger)^{\otimes S}\right)^2.
    \end{align*}
    Next, we introduce arbitrary non-zero numbers $w_S>0$ for each $0\neq S\in[2]^c$ and apply Cauchy-Schwarz in a particular way to find that for any probability distribution $\mu$:
        \begin{align*}
        &\max_{M}\EE_{\vec W\sim\mu}\left[\sum_{a\in A}\frac{1}{\tr[M_a] d^{cn}}\left(\sum_{0\neq S\in[2]^c}\frac{\sqrt{w_S}}{\sqrt{w_S}}(3\veps)^{\abs{S}}\tr[M_a (\vec W+\vec W^\dagger)^{\otimes S}]\right)^2\right]\\
        \leq&\max_M \EE_{\vec W\sim\mu}\left[\sum_{a\in A}\frac{1}{\tr(M_a) d^{cn}}\left(\sum_{0\neq S\in[2]^c}\frac{1}{w_S}(3\veps)^{2\abs{S}}\tr[M_a(\vec W+\vec W^\dagger)^{\otimes S}]^2\right)\left(\sum_{0\neq S\in[2]^c}w_S\right)\right]\\
        \leq&\left(\sum_{0\neq S\in[2]^c}w_S\right)\left(\sum_{0\neq S\in[2]^c}\frac{(3\veps)^{2\abs{S}}}{w_S}\max_M\EE_{\vec W\sim\mu}\left[\sum_a\frac{\tr[M_a(\vec W+ \vec W^\dagger)^{\otimes S}]^2}{d^{cn}\tr(M_a)}\right]\right)\\
        =&\left(\sum_{0\neq S\in[2]^c}(3\veps)^{\abs{S}}\sqrt{\max_M\EE_{\vec W\sim\mu}\left[\sum_a\frac{\tr(M_a(\vec W+\vec W^\dagger)^{\otimes S})}{d^{cn}\tr[M_a]}\right]}\right)^2,
    \end{align*}
    where we picked $w_S=(3\veps)^{\abs{S}}\sqrt{\max_M\EE_\mu\sum_a\frac{\tr(M_a(\vec W+ \vec W^\dagger)^{\otimes S})}{d^{cn}\tr[M_a]}}>0$ to arrive at the last term.
     For any distribution $\mu $ and (finite) \ac{POVM} $M=\{M_a\}_{a\in A}$ we can therefore upper-bound
    \begin{equation}\label{eq:auxiliary_bound}
        \delta_{c,\veps}\leq \left(\sum_{0\neq S\in[2]^c}(3\veps)^{\abs{S}}\sqrt{\EE_{\vec W\sim\mu}\left[\sum_{a\in A}\frac{\tr[M_a (\vec{W}+\vec{W}^\dagger){\otimes S}]^2}{d^{cn}\tr[M_a]}\right]}\right)^2.
    \end{equation} 
In order to expand the tensor product $(\vec W+ \vec W^\dagger)^{\otimes S}$ we introduce the following index set: For a fixed $S\in[2]^c$, let $\mathcal{I}(S)=\{I\in\{-1,0,1\}^c|\supp(S)=\supp(I)\}$. Note that $\abs{\mathcal{I}(S)}=2^{\abs{S}}$, where $\abs{S}$ here refers to the number of non-zero entries of $S$. Due to unitarity, it then holds that $(\vec W+ \vec W^\dagger)^{\otimes S}=\sum_{\tau \in\mathcal{I}(S)}\vec W^{\otimes \tau}$. Via simple algebraic manipulations, and the fact that for any operators $A,B$ where $A$ is positive semidefinite, it holds that $\abs{\tr[AB]}\leq\norm{B}_\infty\tr[A]$, we can derive an upper bound for the term inside the square root in \cref{eq:auxiliary_bound}:
    \begin{align*}
&\EE_{\vec{W}\sim\mu}\left[\sum_{a\in A}\frac{\tr[M_a (\vec{W}+\vec{W}^\dagger)^{\otimes S}]^2}{d^{cn}\tr[M_a]}\right]
=\sum_{a\in A}\frac{1}{d^{cn}\tr[M_a]}\EE_{\vec{W}\sim\mu}\left[\tr\left[M_a\sum_{\tau\in\mathcal{I}(S)} \vec{W}^{\otimes \tau}\right]^2\right]\\
=&\sum_{a\in A}\frac{1}{d^{cn}\tr[M_a]}\EE_{\vec{W}\sim\mu}\left[\tr\left[M_a\otimes M_a\sum_{\tau_1\in\mathcal{I}(S)} \vec{W}^{\otimes \tau_1}\otimes\sum_{\tau_2\in\mathcal{I}(S)} \vec{W}^{\otimes \tau_2}\right]\right]\\
=&\sum_{a\in A}\frac{1}{d^{cn}\tr[M_a]}\sum_{\tau\in\mathcal{I}(S)^2}\tr\left[M_a\otimes M_a \EE_{\vec{W}\sim\mu}[ \vec{W}^{\otimes \tau}]\right]\\
\leq&\sum_{a\in A}\frac{\tr[M_a\otimes M_a]}{d^{cn}\tr[M_a]}\sum_{\tau\in\mathcal{I}(S)^2}\norm{\EE_{\vec{W}\sim\mu} [\vec{W}^{\otimes \tau}]}_\infty
=\sum_{\tau\in\mathcal{I}(S)^2}\norm{\EE_{\vec{W}\sim\mu}[\vec{W}^{\otimes \tau}]}_\infty.\\
\end{align*}
Clearly, for the sites $k$ where $\tau_k=0$ we have $W^{\tau(k)}=\openone$ on this sites for all $\vec W$. Therefore, the sites can be factored into a tensor product of the identity with the sum over the remaining sites. Since tensoring with the identity does not change the operator norm of any operator, we can therefore assume that the strings $\tau$ only take values in $\{\pm1\}$ and are of length $2\abs{S}$. Moreover, for a product distribution $\mu=\nu^{\otimes n}$ with respect to some distribution $\nu$ over $\WW^{(d)}$, we then obtain an upper bound \\
\begin{equation}
\sum_{\tau\in\mathcal{I}(S)^2}\norm{\EE_{\vec W\sim\nu^{\otimes n}}[\vec W^{\otimes\tau}]}_\infty\leq 4^{\abs{S}}\max_{\tau\in\{\pm 1\}^{2\abs{S}}}\norm{\EE_{W\sim\nu}[W^{\otimes\tau}]}_\infty^n.
    \end{equation}
Inserting this upper bound into \cref{eq:auxiliary_bound} yields:

    \begin{align}\label{eq:reduction_to_twirls}
        \delta_{\veps,c}&\leq\left(\sum_{0\neq S\in[2]^c}(6\veps)^{\abs{S}}\sqrt{\max_{\tau\in\{\pm 1\}^{2\abs{S}}}\norm{\EE_{W\sim\mu}[W^{\otimes\tau}]}_\infty}\right)^2\\
        &\leq\max_{1\leq m\leq c}\max_{\tau\in\{\pm 1\}^{2m}}\norm{\EE_{W\sim\mu}[W^{\otimes\tau}]}_\infty\left(\sum_{0\neq S\in[2]^c}(6\veps)^{\abs{S}}\right)^2.
    \end{align}
    Moreover, since $ \sum_{0\neq S\in[2]^c}(6\veps)^{\abs{S}}=(1+6\veps)^c-1\leq 6c\veps e^{6c\veps}$, the claim follows.
     \end{proof}

We continue with deriving the norm bounds on the expression $\norm{\EE_{W\sim\nu}[W^{\otimes\tau}]}_\infty$. For this purpose, according to \cref{lemma:first_delta_bound}, we may pick any distribution $\nu$ over $\WW^{(d)}$ which is convenient for us to work with. Here, we simply use the uniform distribution over $\WW^{(d)}$ when $d=2$, and the uniform distribution over the subset $\hat{\WW}^{(d)}:=\WW^{(d)}\setminus \WW^{(d)}_\infty$ for all other prime-power dimensions. Note, that $\abs{\hat\WW^{(d),n}}=d(d-1)$.

\begin{proposition}\label{prop:prime_power_result}
    Let $d=p^r$ be an odd prime power, $m$ a natural number, $\tau\in\{\pm1\}^{2m}$ and let $M_\tau:=\sum_{W\in\hat{\WW}^{(d)}}W^{\otimes\tau}$. 
    Then we have the following bounds, depending on $m$ and $r$:
    \begin{itemize}
        \item $\norm{M_\tau}_\infty\leq d(d-1)$ when $d\mid m$,
        \item $\norm{M_\tau}_\infty\leq d$ when $d\nmid m$ and $r=1$,
        \item $\norm{M_\tau}_\infty\leq 2dp^{r-1}$ when $d\nmid m$ and $r\geq 2$.
    \end{itemize}
    In particular, with $\nu$ being the uniform distribution over $\hat\WW^{(d)}$ we have that $\norm{\EE_{W\sim\nu}[W^{\otimes\tau}]}\leq 1$ for all $\tau$, and $\norm{\EE_{W\sim\nu}[W^{\otimes\tau}]}\leq\frac{d}{d(d-1)}= \frac{1}{d-1}$, $\norm{\EE_{W\sim\nu}[W^{\otimes\tau}]}\leq \frac{2dp^{r-1}}{d(d-1)}=\frac{2p^{r-1}}{p^r-1}$ for $m<d$ when $r=1,r>1$ respectively.
\end{proposition}
\begin{proof}
While the expression to be estimated looks deceptively like it can be handled by means of representation theory, these methods are not applicable here, as the map $W\mapsto W^{\otimes\tau}$ is \emph{not} a representation due to the occurrence of the inverses. Instead, we will prove the claim by an explicit calculation.
In order to avoid confusion from the start, all occurrences of $\delta$ in this proof refer to the Kronecker delta, not to \cref{eq:delta_quantity}. Also, for simplicity, we assume $m\leq d$, since only the parity of $m$ modulo $d$ is important here.
In order to keep the presentation of the proof as neat as possible, let $\kappa :=\langle \tau,\mathbf{1}\rangle/2 \in\ZZ$ and $T(J):=\langle \tau,J\rangle$ for $J\in[d]^{2m}$.  As we will show in the following, we can express $M_\tau$ as
    \begin{equation}\label{eq:average_operator_solution_set_form}
        M_\tau=\sum_{W\in\hat{\WW}} W^{\otimes \tau}=d\sum_L\sum_{0\neq\Delta\in\mathfrak{S}(L)}\ket{L+\Delta\tau}\bra{L},
    \end{equation}
    where $\mathfrak{S}(L)$ are the solutions to some quadratic equation, depending on $L$ (also $m,\kappa$, but these we consider to be fixed throughout). From here we can then estimate the norm, where we exploit the fact that the norm of \cref{eq:average_operator_solution_set_form} is upper bounded by the same expression when, instead of $\mathfrak{S}(L)$, we use some superset thereof in the summation. First, however, we derive the expression \cref{eq:average_operator_solution_set_form}.
We may assume \obda that $\tau_1=1$, since exchanging $W\leftrightarrow W^{-1}=W^\dagger$ in the summation leads to the same expression. We split the sum into summations over the subgroups $\mathcal{W}_a=\langle W_a\rangle $ for $a=0,\dots,d-1$, (which only intersect at the identity) and express the operators in terms of their eigenprojectors.
    \begin{align*}
        \sum_{W\in\hat{\WW}} W^{\otimes \tau}&=\sum_{\alpha\in[d]}\sum_{k\in[d]}(W_\alpha^k)^{\otimes \tau}-d\mathds{1}=\sum_{\alpha}\sum_{k}\sum_{J\in[d]^{2m}} \omega^{k\langle\tau,J\rangle}\psi^\alpha_J-d\mathds{1}\\&=d\sum_{\alpha}\sum_{\langle\tau,J\rangle=0}\psi^\alpha_J-d\mathds{1}.
    \end{align*}
    For the next step, we exploit the fact that the $\omega^j$-eigenstates of the different $W_a$ are related to each other via the action of the $S$-gate, and spell out everything in the computational basis. Also, we abbreviate $f(K)=\sum_i k_i(k_i-1)/2=\langle K,K-\mathbf{1}\rangle/2$, which is always an integer. We know expand the $\ket{\psi^\alpha_j}$ in terms of the computational basis and find
\begin{equation}
    \psi^\alpha_J=\sum_{K,L\in[d]^{2m}}\omega^{\avg{J,L}-\avg{J,K}}\omega^{\alpha (f(K)-f(L)}\kb{K}{L}
\end{equation}
    
    \begin{align*}    \sum_\alpha\sum_{T(J)=0}\psi^\alpha_J
        &=d^{-2m}\sum_\alpha\sum_{T(J)=0}\sum_{K,L}\omega^{\avg{J,L}-\avg{J,K}}\omega^{\alpha (f(K)-f(L)}\kb{K}{L}\\
        &=d^{-1}\sum_\alpha\sum_{K,L}\delta_{(l_1-k_1)\tau=_dL-K}\omega^{\alpha (f(K)-f(L)}\kb{K}{L} \\
        &=\sum_{K,L}\delta_{\Delta\tau=_dK-L}\delta_{f(K)=_df(L)}\kb{K}{L}=\sum_L\sum_{\Delta\in\mathfrak{S}(L)}\kb{L+\Delta\tau}{L}.
    \end{align*}
    In the second line we applied \cref{lemma:linear_constrained_fourier_sum}, and in the third line we substituted $\Delta:=k_1-l_1$, and let $\mathfrak{S}(L)=\{\Delta:f(L+\Delta\tau)=f(L)\mod d\}$. That is in order to determine $M_\tau$ we need to determine the solutions $\mathfrak{S}(L)$.
    \begin{align}
        \nonumber &f(L+\Delta\tau)-f(L)=0\mod d\\\Leftrightarrow&\left(\langle L+\Delta\tau,L+\Delta\tau\rangle-\langle L+\Delta\tau,\mathbf{1}\rangle-\langle L,L\rangle+\langle L,\mathbf{1}\rangle\right)/2=0\mod d,\\
    \nonumber \Leftrightarrow&\;(2\Delta\langle\tau,L\rangle+\Delta^2\langle\tau,\tau\rangle-\Delta\langle\tau,\mathbf{1}\rangle)/2=0 \mod d,\\
        \nonumber \Leftrightarrow&\;\Delta(m\Delta+ T(L)-\kappa)=0\mod d.
        \end{align}
Until here, the derivation for the cases $r=1$ and $r>1$ is completely analogous. Clearly, $\Delta=0$ is always a solution. However, when $r>1$, $\ZZ_d$ is not a field, and we also need to consider the possibility of zero-divisors to analyze the solutions $\mathfrak{S}(L)$. We begin with the case where $r=1$. 
In this case, apart from $\Delta=0$ we also find the solution $\Delta=m^{-1}(\kappa-T(L))$, which of course only gives an additional solution when $T(L)-\kappa\neq 0$. In particular, we get 
\begin{equation}
    M_\tau = d\sum_{T(L)-\kappa\neq 0}\kb{L+m^{-1}(\kappa-T(L))\tau}{L}.
\end{equation}
In particular, we observe that the mapping $L\mapsto L+m^{-1}(\kappa-T(L))\tau$ is injective:
Let $J,K$ such that 
    \begin{equation}\label{eq:injectivity}
        J+m^{-1}(\kappa-T(J))\tau=K+m^{-1}(\kappa-T(K))\tau.
    \end{equation}
    By applying $T(\cdot)$ to this condition we find that $T(J)=T(K)$ must hold, since $T(\tau)=\langle\tau,\tau\rangle=2m$. Clearly, inserting this relation back into \cref{eq:injectivity} it then also follows that $J=K$.\\

For $r>1$, we now additionally need to account for solutions arising from non-trivial zero-divisors. However, we do not actually consider these solutions in more detail, and just use that these must satisfy $\Delta\subset p\ZZ$. Note that there exist configurations of $m<d$ and $L$ for which indeed $\mathfrak{S}_{\tor}(L)=p\ZZ$ holds, such that this assumption is not unnecessarily pessimistic. Thus it remains for us to identify the additional solutions arising from  $m\Delta+T(L)-\kappa=0$. Let $t,s$ be the maximal integer such that $p^t|(T(L)-\kappa)$ and $p^s|m$. If $t<s$, then this equation cannot have any solutions, since $p$ (and powers thereof) does not have a multiplicative inverse. On the other hand, when $s<t$, all solutions must be contained within $p\ZZ$ for the same reason. Hence, we only obtain additional solutions in $\mathfrak{S}(L)$ not already contained in $p\ZZ$ only for $s=t$. In this case, let $\frac{p^s}{m}$ denote the formal inverse of $\frac{m}{p^s}$ modulo $d$. Then, we have that all $\Delta = \frac{p^s}{m}\frac{\kappa-T(L)}{p^s}+ap^{r-s},a=0,\dots,p^{s}-1$ are additional solutions. Importantly, these are also the only solutions, since any two $\Delta,\Delta'$ satisfying $m\Delta+T(L)-\kappa=m\Delta'+T(L)-\kappa=0$ must differ by a zero-divisor for $m$, and hence are contained in $p\ZZ$. Therefore, we find that
\begin{equation}
    \norm{M_\tau}\leq d\sum_{a=1}^{p^{r-1}}\left(\norm{\sum_L\kb{L+a p\,\tau}{L}}_\infty+\norm{\sum_L\kb{L+\left(\frac{p^s}{m}\frac{\kappa-T(L)}{p^s} + ap\right)\tau}{L}}_\infty\right).
\end{equation}
Now observe, that like in the case $r=1$, each of the functions of $L$ in the $\ket{\cdot}$ above is injective, since $a$ is fixed. In particular, all the operators in the sum over $a$ have norm one, and hence
\begin{equation}
    \norm{M_\tau}_\infty\leq 2dp^{r-1}=2\frac{d^2}{p},
\end{equation}
which finishes the proof.

\end{proof}

\begin{techlemma}\label{lemma:qubit_Pauli_twirling}
    Let $k$  be an integer and $\mathcal{P}=\{I,X,Y,Z\}$ the set of the four Pauli matrices. Then:
    \begin{equation}
        \norm{\sum_{P\in\mathcal{P}}P^{\otimes 2k}}_\infty = 3 + (-1)^k.
    \end{equation}
\end{techlemma}
\begin{proof}
    This is a straightforward calculation: We substitute $X = \kb{0}{1}+\kb{1}{0}$, $Y = i\kb{0}{1}-i\kb{1}{0}$, and $I,Z$ analogously. Then
    \begin{align}
        \sum_{P\in\{I,X,Y,Z\}}P^{\otimes 2k} &= \sum_{I\in[2]^{2k}}\kb{I}{I} + (-1)^{\abs{I}}\kb{I}{I} + \kb{I}{I\oplus\mathbf{1}}+i^{2k}(-1)^{\abs{I}}\kb{I}{I\oplus\mathbf{1}}\\
        &=\sum_{I\in[2]^{2k}}(1+(-1)^{\abs{I}})\kb{I}{I} + (1+(-1)^{k+\abs{I}})\kb{I}{I\oplus\mathbf{1}}.
    \end{align}
    In particular, when $k$ is odd, for every $I\in[2]^k$ either of the two factors will vanish, leading to $ \norm{\sum_{P\in\mathcal{P}}P^{\otimes 2k}}_\infty=2$. On the other hand, when $k$ is even, with $\ket{\GHZ_n}$ being the GHZ-state on $n$ qubits, we clearly find $\sum_{P\in\mathcal{P}}P^{\otimes 2k}\ket{\GHZ_{2k}}=4\ket{\GHZ_{2k}}$, which concludes the proof. 
\end{proof}

\begin{repproposition}{prop:composite_norm_bounds}[restated]
    Let $d\geq 2$ be any integer with prime factorization $d=p_1^{r_1}\dots p_l^{r_l}$ in increasing order $p_1<\dots<p_l$, and $\tau\in\{\pm1\}^{2m}$ with $m<d$. Then there exists a distribution $\mu$ over $\WW^{(d)}$ such that  
    \begin{equation}
        \norm{\EE_{W\sim\mu}[W^{\otimes\tau}]}\leq\max_j \chi(p_j^{r_j}),\text{ where }\chi(p^{r}):=\begin{cases}
            \frac{1}{2},&p=2\text{ and }r=1\\
            \frac{1}{p-1}, &p\neq 2\text{ and }r=1\\
            \frac{2p^{r-1}}{p^r-1} &\text{ else}.
        \end{cases}
    \end{equation}
    In particular, if $d$ is not divisible by four, then we find $\max_j \chi(p_j^{r_j})\leq\frac{3}{4}<1$.
\end{repproposition}

\begin{proof}
    First of all, by a variant of the Chinese Remainder Theorem (CRT), we find that, up to modification of phases and local unitary Clifford equivalence, we have that $\WW^{(d)}\cong \WW^{(p_1^{r_1})}\otimes\dots\otimes\WW^{(p_l^{r_l})}$, see e.g. \cite{Ellinas1999Prime}. Specifically, one can consider the map \[X_p^{\alpha_1}Z_p^{\beta_1}\otimes X_q^{\alpha_2}Z_q^{\beta_2}\mapsto X_{pq}^{\alpha_1q+\alpha_2 p }Z_{pq}^{\beta_1 q+\beta_2 p},\] 
    which, according to the CRT, is indeed bijective when $p,q$ are relatively prime.
    Now, let $\nu_j$ denote the uniform distribution over $\hat{\WW}^{(p_j^{r_j})}$ for $p_j\neq 2$ or the uniform distribution over $\WW^{(p_j)}$ when $p_j=2$ and $\nu=\nu_1\otimes\dots\otimes\nu_l $ denote the product distribution of these. Then for any $m<d$ and $\tau\in\{\pm1\}^n$ we have
    \begin{equation}
        \norm{\EE_{W\sim\nu}[W^{\otimes\tau}]}_\infty=\prod_{j=1}^l\norm{\EE_{W_j\sim\nu_j}[W_j^{\otimes\tau}]}_\infty.
    \end{equation}
    Now, according to \cref{prop:prime_power_result}, and \cref{lemma:qubit_Pauli_twirling} we have $\norm{\EE_{W_j\sim\nu_j}[W_j^{\otimes \tau}]}\leq 1$ in any case. Moreover, since $m<d$ by assumption, there is at least one $p\in\{p_1,\dots,p_l\}$ which does not divide $m$. With $r$ denoting the corresponding power of $p$ in the prime factorization of $d$, we get $\norm{\EE_{W\sim\nu}[W^{\otimes\tau}]}_\infty\leq\frac{1}{2}$ when $r=1$ and $p=2$; $\norm{\EE_{W\sim\nu}[W^{\otimes\tau}]}_\infty\leq\frac{1}{p-1}$ when $r=1$ and $p\geq 3$; and $\norm{\EE_{W\sim\nu}[W^{\otimes\tau}]}_\infty\leq\frac{2p}{p^{r}-1}$ in all other cases, as evident from \cref{prop:prime_power_result} and \cref{lemma:qubit_Pauli_twirling}. Picking the maximum among these values for all prime factors $p_j$ accordingly clearly results in a valid upper bound for all $m<d$, which proves the claim. 
\end{proof}

\begin{corollary}
    For $4\nmid d$ and $d\geq 2$, and $\veps<\frac{1}{6}$, the task of Weyl-Heisenberg amplitude estimation is of degree $d$.
\end{corollary}

Finally, to conclude this section, we provide the proof that the mimicking state algorithm \cref{alg:protocol} works correctly.
\begin{techlemma}\label{lemma:correctness}
    Let $\veps>0$ and $\rho$ a fixed quantum states, and $u_W$ estimates of $ \abs{\tr(\vec W\rho)}$ up to additive error $\veps/6$. Then with probability $p\geq 0.99$ the output of \cref{alg:protocol} provides a classical description of a state $\sigma$ such that $\WW_{\veps,\rho}\subset\WW_{\veps/3,\sigma}$.
\end{techlemma}
\begin{proof}
    First of all, the algorithm is guaranteed to be correct (with high probability) whenever no violating $W\in\WW^n$ is found in step 1.\\
    Let us therefore assume that the algorithm iterates over all $t=0,\dots,T$ and reaches the last step $t=T$. Let $y_t=\tr(\hat{W}_t\sigma_t)$ for $0\leq t\leq T$. By Lemma 3.1 from \cite{king2025triply} we are guaranteed that after $T$ steps of the algorithm: 
    \begin{align*}
        \sum_{t=1}^T\sgn(y_t-\hat{z}_t)\cdot y_t&\leq\lambda_{\min}\left(\sum_{t=1}^T\sgn(y_t-\hat{z}_t)\cdot\hat{\vec W}_t\right)+2\sqrt{nT}\\
        &\leq\sum_{t=1}^T\sgn(y_t-\hat{z}_t)\tr(\hat{\vec W}_t\rho)+2\sqrt{nT}.
    \end{align*}
    This can be equivalently expressed as
    \begin{equation}\label{eq:algorithm_key_estimate}
        \sum_{t=1}^T\sgn(y_t-\hat{z}_t)\left(y_t-\tr(\hat{\vec W}_t\rho)\right)\leq2\sqrt{nT}.
    \end{equation}
    Next, observe that we can lower bound the difference 
    \begin{align*}
        \abs{\tr(\vec W_t\sigma_t)-z_t}\geq \abs{\abs{\tr(\vec W_t\sigma_t)}-u_t}-\abs{\abs{z_t}-u_t}\geq \veps/2-(\veps/8+\veps/30)
>\veps/3.
\end{align*}
This in turn leads to the following estimates:
    \begin{align}
        \abs{y_t-\hat{z}_t}=\abs{\tr(\hat{\vec W}_t\sigma_t)-\hat{z}_t}\geq\frac{1}{\sqrt{2}}\abs{\tr(\vec W_t\sigma_t)-z_t}\geq\frac{\veps}{3\sqrt{2}},\\
    \label{eq:algorithm_lower_bound1}\abs{y_t-\tr(\hat{\vec W}_t\rho)}\geq\abs{y_t-\hat{z}_t}-\abs{\hat{z}_t-\tr[\hat{\vec W}\rho]}\geq\frac{\veps}{3\sqrt{2}}-\frac{\veps}{30}>\frac{\veps}{5}.
\end{align}
The first estimate, together with $\abs{z_t-\tr(W_t\rho)}\leq\veps/30$ can be used to show that $\sgn(y_t-\hat{z}_t)=\sgn(y_t-\tr(\hat{W}_t\rho))$. This can be easily shown by distinguishing the following two cases:
\begin{align*}
   \text{If  }\;y_t-\hat{z}_t\geq\frac{\veps}{3\sqrt{2}}\;\Rightarrow\;y_t-\hat{z}_t-(\tr(\hat{\vec W}_t\rho)-\hat{z}_t)\geq\frac{\veps}{3\sqrt{2}}-\frac{\veps}{30}>0,\\
   \text{If  }\;\hat{z}_t-y_t\geq\frac{\veps}{3\sqrt{2}}\;\Rightarrow\;\hat{z}_t-y_t-(\hat{z}_t-\tr(\hat{\vec W}_t\rho))\geq\frac{\veps}{3\sqrt{2}}-\frac{\veps}{30}>0.
\end{align*}
We use this fact, together with \cref{eq:algorithm_lower_bound1} to lower bound the left hand side of \cref{eq:algorithm_key_estimate}:
\begin{align*}
        &2\sqrt{nT}\geq \sum_{t=1}^T\sgn(y_t-\hat{z}_t)\left(y_t-\tr(\hat{\vec W}_t\rho)\right)=\sum_{t=1}^T\abs{y_t-\tr(\hat{\vec W}_t\rho)}\geq \frac{\veps}{5}T,\\
        \Leftrightarrow \;\;&T\leq 100 n/\veps^2.
    \end{align*}
This, however, is a contradiction to the way $T$ was chosen beforehand, i.e., the algorithm never reaches the last step.
Let us finally estimate the probability that this algorithm fails. The only place where it can lead to erroneous results is when estimating the $z_t$ in step 3a). The probability that all of these estimates are correct is $p\geq (1-0.01/\abs{T})^{\abs{T}}\geq 1-0.01$. Hence, it follows that the algorithm outputs a correct mimicking state $\sigma$ in $99$ out of $100$ runs. 
\end{proof}

\section{Remarks on Numerical Simulations}\label{sec:numerics}

To quantify the sampling requirements of the proposed certification protocol, we performed numerical simulations for systems of up to $10$ qutrits. For each system size, we determine the minimal number of samples $N_{\min}$ that guarantees a success probability $p_{\mathrm{succ}} \ge 0.7$. 
The success probability measures how often the protocol correctly identifies all tested expectation values $\operatorname{tr}(W\rho)$ within some tolerance.

\subsection{Wilson confidence strategy}\label{subsec:Wilson}

To decide whether a given number of samples is sufficient, we employ the Wilson confidence-interval strategy for binomial success probabilities. Given $t$ repetitions of the reconstruction experiment at fixed sample size $N$ with $s$ observed successes, the estimated probability is $\hat{p}=s/t$. The Wilson interval \cite{brown2001interval} provides an adjusted confidence bound that remains reliable even for small $t$, 
\begin{align}
\Pr\left[ p \in  (w_+, w_-)\right] = 1- \alpha \quad\text{with } w_{\pm} =
\frac{n\hat{p} + \frac{z^2}{2}}{t + z^2}
\pm 
\frac{z\sqrt{t}}{t+z^2}
\sqrt{\hat{p}(1-\hat{p}) + \frac{z^2}{4t}},
\end{align}
where $z$ is the standard normal quantile corresponding to the desired confidence level $1-\alpha=90\%$.  
The lower bound of this interval serves as a conservative estimate of the true success probability. 
For a target success probability $p_\star=0.7$, we certify that the true success rate at sample size $N$ satisfies $p_N \ge p_\star$ whenever $w_{-}\ge p_\star$.
At each fixed $N$, we apply the following stopping rule:
\begin{align}    
\text{Decision} =
\begin{cases}
\textbf{Accept}, & \text{if } w_- \ge p_\star, \\[4pt]
\textbf{Early-reject}, & \text{if } w_+ < p_\star, \\[4pt]
\text{Continue sampling (increase $t$) up to } t_{\max}, & \text{otherwise.}
\end{cases}
\end{align}

If $t$ reaches $t_{\max}$ while $w_- < p_\star \le w_+$ (undecided), we conservatively treat this $N$ as a reject.

Across sample sizes, we perform an exponential search: starting at $N_0$, increase $N \leftarrow \max\{N+1,\,\lfloor gN\rfloor\}$ (with growth factor $g>1$) whenever the decision at the current $N$ is reject or undecided, until we encounter the first accepted $N$ (call it $N_{\mathrm{hit}}$). If a last failing point $N_{\mathrm{fail}}$ is known, we optionally refine via bisection on the integer interval $(N_{\mathrm{fail}},\,N_{\mathrm{hit}}]$ using the same Wilson decision at each midpoint, returning the minimal accepted $N$. This value is reported as $N_{\min}$.

\subsection{Three-copy strategies}

In the three-copy setting we estimate, for an $n$-qutrit state $\rho$, the expectations $\langle\vec W\rangle=\mathrm{tr}(\vec W\rho)$ of Weyl strings $\vec W\in\WW^{\otimes n}$, where $\WW$ denotes the single-qutrit Weyl–Heisenberg set (of size $|\WW|=d^2=9$). All $|\WW|^{\,n}=9^{\,n}$ strings are verified simultaneously from the same 3-copy data.
A reconstruction experiment counts as a success if, for every $\vec W\in\WW^{\otimes n}$, the reconstructed value $\hat{y}$ sufficiently matches the theoretical target. That is $\max_\vec{W} |\tr( \vec W^{\otimes 3} \rho^{\otimes 3}) - \hat{y}| < \delta$ where we used $\delta=0.1$ in our experiments. At fixed sample size $N$, the accept/reject decision uses the Wilson rule of Sec.~\ref{subsec:Wilson} to certify $p_{\mathrm{succ}}\ge 0.7$; the minimal accepted $N$ is reported as $N_{\min}$.

\subsection{Single-copy strategies}

In the single-copy baseline we fix a hidden Weyl string $\vec W^\star\in\WW^{\otimes n}$ (which determines $\rho$) and draw measurement strings $\vec W$ uniformly at random from $\WW^{\otimes n} \;=\; \{\,X^{a_1}Z^{b_1}\otimes\cdots\otimes X^{a_n}Z^{b_n}\;:\; a_j,b_j\in\{1,2\}\,\}$,
so that $|\WW^{\otimes n}|=4^n$ (each site uses one of the four non-identity pairs $(a,b)\in\{1,2\}^2$). A draw is a success iff $\vec W=\vec W^\star$ or $\vec W={\vec W^\star}^\dagger$. Hence, a single draw has hit probability $p_{\text{hit}}=2/4^{-n}$ and, with $N$ independent draws, the probability of at least one hit is:
\begin{align}
p(N)=1-(1-2/4^{-n})^{N}. \label{eq:hit_prob_single_copy}
\end{align}

For each fixed $N$, we repeat $t=2000$ times and record the number of successful retrievals. The empirical success probability is $\hat p(N)=s/t$. We iterate over a fine grid of $N$ and report 
\begin{align}
N_{\min}^{\mathrm{emp}}=\min\{N:\hat p(N)\ge 0.7\}.
\end{align}

\subsection{Simulation summary}

Figure~\ref{fig:3c_vs_1c} summarizes the scaling of the minimal sample number $N_{\min}$ with the number of qutrits. The five curves correspond to the strategies described above: three-copy full check, three-copy random check (Wilson), three-copy random check (empirical), one-copy empirical, and one-copy theoretical. The data show that the three-copy protocol dramatically reduces the sampling complexity compared to the single-copy baseline to achieve the prescribed success probability.

\section{Technical Lemmata}\label{sec:lemmata}

\begin{techlemma}[Hoeffding]\label{lemma:Hoeffding}
    Let $X_1,\dots,X_N$ be independent real-valued random variables, not necessarily identically distributed, such that $a_i\leq X_i\leq b_i$ for suitable $a_i,b_i$ almost surely. Then, for any $c\geq 0$ it holds that
    \begin{equation}
        \Pr\left[\abs{\frac{1}{N}\sum_{i=1}^N X_i-\EE[X_i]}\geq c\right]\leq 2\exp\left(-\frac{2c^2N^2}{\sum_i(b_i-a_i)^2}\right).
    \end{equation}
\end{techlemma}
\begin{techlemma}\label{lemma:linear_constrained_fourier_sum}
    Let $d$ be a prime, $\omega $ a primitive $d-th$ root of unity, and $\mathbf{a},\mathbf{b}\neq 0$ be vectors on $\ZZ_d^m$ and assume that $a_1=1$. Then
    \begin{equation}
        \sum_{\langle \mathbf{a},I\rangle=0}\omega^{\langle \mathbf{b},I\rangle}=d^{m-1} \delta_{b_1\mathbf{a}=\mathbf{b}}.
    \end{equation}
\end{techlemma}
\begin{proof}
First of all, the condition $\langle a,I\rangle=0$ translates to $i_1=-\sum_{k\geq 2}i_k a_k$. Therefore, we express the summation as
\begin{align*}
    \sum_{i_2,\dots,i_m}\omega^{-b_1\sum_{k\geq 2} a_ki_k}\prod_{k\geq 2}\omega^{b_k i_k}=\sum_{i_2,\dots,i_m}\prod_{k\geq 2}\omega^{i_k(b_k-b_1 a_k)}=d^{m-1}\prod_{k\geq 2}\delta_{b_k=b_1a}=d^{m-1}\delta_{b_1\mathbf{a}=\mathbf{b}},
\end{align*}
where we used the fact that the condition $b_1a_1=b_1$ is trivially satisfied by construction in the last step.
\end{proof}

\section{Quantum Circuits}\label{sec:q_circuits}
In this section we provide examples of quantum circuits to implement the measurement processes in the main part.
\subsection{Quantum circuits for generalized Bell-basis measurements}
In order to show that the quantum circuits required to execute the generalized Bell-basis measurements are efficient, we sketch their structure, which strongly resembles the circuits for the preparation of ordinary Bell states in the circuits below. Specifically, for $n=1$, the diagonalizing unitary can be decomposed into single- and two-qubit qudit Cliffords as
\begin{equation}
C = H^{(1)}\circ CX^{d-1,d}\circ CX^{d-2,d-1}\circ\dots\circ CX^{1,2}.
\end{equation}
For general $n$, the diagonalizing unitary is then simply $C^{\otimes n}$, which is schematically depicted in the circuit below for the example of qu\emph{trits}.

\begin{equation}\label{circ:qutrit_measurement}
\begin{split}
\begin{quantikz}[row sep=0.15cm, column sep=0.3cm]
\lstick[4]{$\rho$} &  &         &&  & \ctrl{5} && \gate{F_3} &  \\
                   &  &   &&  \ctrl{5} &        && \gate{F_3} &  \\
\setwiretype{n}    & \myvdots & & \myidots &     &&           &            &     \\
                   & &\ctrl{5} &    &  &        && \gate{F_3} &  \\
\\
\lstick[4]{$\rho$} &&  &         &  & \gate{X_3} & \ctrl{5}  &&  \\
                   &&  &  & \gate{X_3} & \ctrl{5}   &        &&  \\
\setwiretype{n}    & \myvdots &   & \myidots & \myidots &     &            &&     \\
                   && \gate{X_3} & \ctrl{5} &  &         &        &&  \\
\\
\lstick[4]{$\rho$} &&  &         &  &         & \gate{X_3} &&  \\
                   &&  &         &  & \gate{X_3} &         &&  \\
\setwiretype{n}    & \myvdots &    &            & \myidots &&      &            &     \\
                   &&  & \gate{X_3} &  &         &         &&  \\
\end{quantikz}
\end{split}
\end{equation}

Here, the notation of the ordinary controlled X gate refers to the two-qudit unitary defined in \cref{eq:CX-gate_Def}, and $F_3$ is the Fourier gate as in \cref{eq:H-gate_Def}.
\subsection{\texorpdfstring{Transpilation to qubit architectures for the special case $d=3$}{Transpilation to qubit architectures for the special case d=3}}

Clearly, in order to faithfully represent qutrit gates, it suffices to embed one qutrit into two qubits. Here, we have to ensure that the representation is reducible (into a trivial one-dimensional component and a three-dimensional irreducible one). Specifically, we choose the embedding via the binary representation, i.e. $\ket{0}\mapsto\ket{00},\ket{1}\mapsto\ket{01},\ket{2}\mapsto\ket{10}$. First, we consider the qutrit Fourier gate $F_3$. The eigenvalues of the Blockmatrix $\begin{pmatrix}
    F_3 & 0\\
    0 & 1
\end{pmatrix}$
are given by $\{1,i,-1,1\}$ with the corresponding (unnormalized) eigenvectors $\{(1-\sqrt{3},1,1,0),(0,1,-1,0),(1+\sqrt{3},1,1,0),(0,0,0,1)\}$. Alltogether, we can therefore can express it as  

\begin{equation}\label{eq:H_circuit}
    \begin{quantikz}[column sep=0.3cm, row sep=0.2cm]
    & \gate[wires=2]{F_3} & \ghost{-S^\dagger} \\
    & \ghost{F_3}         & \ghost{R_y^\dagger(\theta)}
    \end{quantikz}
=
    \begin{quantikz}[column sep=0.3cm, row sep=0.2cm]
    & \ctrl{1} & \gate{HZ} & \octrl{1} & \gate{-S^\dagger} & \octrl{1} & \gate{ZH} & \ctrl{1} & \\
    & \targ{} & \ctrl{-1} & \gate{R_y^\dagger(\theta)} & \ctrl{-1} & \gate{R_y(\theta)} & \ctrl{-1} & \targ{} &
    \end{quantikz}
\end{equation}

where $\theta = \arccos(1/\sqrt{3})$ and $H=(X+Z)/\sqrt{2},S=\kb{0}{0}+\ii\kb{1}{1}$ are the standard qubit Hadamard and $S$-gate and $R_y(\theta)=\cos(\theta)\mathds1+\ii\sin(\theta)Y$. Let us continue with a suitable implementation of the $X_3$-gate before moving on to its controlled version. It is straightforward to verify the following circuit identity:
\begin{equation}\label{eq:X_circuit}
\begin{quantikz}[column sep=0.3cm, row sep=0.2cm]
& \gate[wires=2]{X_3} & \\
& \ghost{X}&
\end{quantikz}
=
\begin{quantikz}[column sep=0.3cm, row sep=0.2cm]
& \octrl{1} & \targ{}       & \ghost{X} \\
& \targ{}        & \octrl{-1}& \ghost{X}
\end{quantikz}
\end{equation}

Since we are working with ternary logic, we know that $-k = 2k$, and we can realize the controlled-X operation conveniently as follows:
\begin{equation}\label{eq:CX_circuit}
\begin{quantikz}[column sep=0.3cm, row sep=0.2cm]
& \gate[wires=4]{CX_3} &\ghost{X_3^\dagger} \\
& \ghost{X_3^\dagger} & \\
& \ghost{X_3^\dagger} & \\
& \ghost{X_3^\dagger} & 
\end{quantikz}
\quad=\quad
\begin{quantikz}[column sep=0.3cm, row sep=0.2cm]
& \ghost{X_3^\dagger} & \ctrl{2} & \\
& \ctrl{1} & \ghost{X_3^\dagger} & \\
& \gate[wires=2]{X_3} & \gate[wires=2]{X_3^\dagger} &\\
& \ghost{CX_3} & \ghost{CX_3} &
\end{quantikz}
\quad=\quad
\begin{quantikz}[column sep=0.3cm, row sep=0.2cm]
&           &            & \ctrl{2}     & \ctrl{2}  & & \ghost{X_3^\dagger}\\
& \ctrl{1}  & \ctrl{1}   &              &           & & \ghost{X_3^\dagger}\\
& \octrl{1} & \targ{}    & \targ{}      & \octrl{1} & & \ghost{X_3^\dagger}\\
& \targ{}   & \octrl{-1} & \octrl{-1}   & \targ{}   & & \ghost{X_3^\dagger}
\end{quantikz}
\end{equation}

\begin{equation}
\begin{quantikz}[column sep=0.3cm, row sep=0.2cm]
& \gate[wires=4]{CX_3} &\ghost{X} \\
& \ghost{X} & \\
& \ghost{X} & \\
& \ghost{X} & 
\end{quantikz}
\quad=\quad
\begin{quantikz}[column sep=0.3cm, row sep=0.2cm]
&  & \ctrl{2} & \\
& \ctrl{1} &  & \\
& \gate[wires=2]{X_3} & \gate[wires=2]{X_3^\dagger} &\\
&  &  &
\end{quantikz}
\quad=\quad
\begin{quantikz}[column sep=0.3cm, row sep=0.2cm]
&           &            & \ctrl{2}     & \ctrl{2}  & & \\
& \ctrl{1}  & \ctrl{1}   &              &           & & \\
& \octrl{1} & \targ{}    & \targ{}      & \octrl{1} & & \\
& \targ{}   & \octrl{-1} & \octrl{-1}   & \targ{}   & & 
\end{quantikz}
\end{equation}

In particular, applying these transpilation rules to the circuit \cref{circ:qutrit_measurement}, we find the following measurement circuit for the $d$-copy efficient protocol on an $n$-qutrit state, where we omit the computational basis measurement sign at the end of each line.
\begin{equation}
\begin{quantikz}[column sep=0.14cm, row sep=0.2cm] 
    \lstick[5]{$\rho$} 
    & & & & & & & & & \ctrl{7} & \ctrl{7} & & & \ctrl{1} & \gate{HZ} & \octrl{1} & \gate{-S^\dagger} & \octrl{1} & \gate{ZH} & \ctrl{1} & \\
    & & & & & & & \ctrl{6} & \ctrl{6} & & & & & \targ{} & \ctrl{-1} & \gate{R_y^\dagger(\theta)} & \ctrl{-1} & \gate{R_y(\theta)} & \ctrl{-1} & \targ{} & \\
    \setwiretype{n} & \myvdots & & & & & \myidots & & & & & & & \myvdots & & & & & & \myvdots & \\ 
    & & & & \ctrl{7} & \ctrl{7} & & & & & & & & \ctrl{1} & \gate{HZ} & \octrl{1} & \gate{-S^\dagger} & \octrl{1} & \gate{ZH} & \ctrl{1} & \\ 
    & & \ctrl{6} & \ctrl{6} & & & & & & & & & & \targ{} & \ctrl{-1} & \gate{R_y^\dagger(\theta)} & \ctrl{-1} & \gate{R_y(\theta)} & \ctrl{-1} & \targ{} & \\ 
    \\
    \lstick[5]{$\rho$} 
    & & & & & & & \octrl{1} & \targ{} & \targ{} & \octrl{1} & & & \ctrl{7} & \ctrl{7} & & & & & & \\ 
    & & & & & & & \targ{} & \octrl{-1} & \octrl{-1} & \targ{} & \ctrl{6} & \ctrl{6} & & & & & & & & \\ 
    \setwiretype{n} & \myvdots & & & & & \myidots & & & & \myidots & & & & & & & & & \\ 
    & & \octrl{1} & \targ{} & \targ{} & \octrl{1} & & & \ctrl{7} & \ctrl{7} & & & & & & & & & & & \\ 
    & & \targ{} & \octrl{-1} & \octrl{-1} & \targ{} & \ctrl{6} & \ctrl{6} & & & & & & & & & & & & & \\ 
    \\[-2pt] 
    \lstick[5]{$\rho$} & & & & & & & & & & & \octrl{1} & \targ{} & \targ{} & \octrl{1} & & & & & & \\ 
    & & & & & & & & & & & \targ{} & \octrl{-1} & \octrl{-1} & \targ{} & & & & & & \\ 
    \setwiretype{n} & \myvdots & & & & & & & & & \myidots & & & & & & & & & & \\ 
    & & & & & & \octrl{1} & \targ{} & \targ{} & \octrl{1} & & & & & & & & & & & \\ 
    & & & & & & \targ{} & \octrl{-1} & \octrl{-1} & \targ{} & & & & & & & & & & &
\end{quantikz}
\end{equation}

\nocite{huang2022quantum,huang2021information,tran2025one,aaronson2020shadow,huang2020predicting,king2024exponential}
\bibliographystyle{unsrt}
\bibliography{references}
\end{document}